\documentclass[prx,superscriptaddress,floatfix,twocolumn]{revtex4-1}
\usepackage{verbatim}
\usepackage{amsmath}
\usepackage{amsfonts}
\usepackage{amssymb}
\usepackage{graphicx}
\usepackage{bm}
\usepackage{color}
\usepackage[active]{srcltx}
\usepackage{cases}
\usepackage{ulem}
\usepackage{multirow}
\usepackage{braket}
\usepackage{colortbl}

\def\stacksymbols #1#2#3#4{\def\theguybelow{#2}
    \def\verticalposition{\lower#3pt}
    \def\spacingwithinsymbol{\baselineskip0pt\lineskip#4pt}
    \mathrel{\mathpalette\intermediary#1}}
\def\intermediary#1#2{\verticalposition\vbox{\spacingwithinsymbol
      \everycr={}\tabskip0pt
      \halign{$\mathsurround0pt#1\hfil##\hfil$\crcr#2\crcr}}}

\begin{document}
%
\title{Topological Quantum Walk with Discrete Time-Glide Symmetry}

\author{Ken Mochizuki}
\affiliation{Department of Applied Physics, Hokkaido University, Sapporo 060-8628, Japan}

\author{Takumi Bessho}
\affiliation{Yukawa Institute for Theoretical Physics, Kyoto University, Kyoto 606-8502, Japan}

\author{Masatoshi Sato}
\affiliation{Yukawa Institute for Theoretical Physics, Kyoto University, Kyoto 606-8502, Japan}

\author{Hideaki Obuse}
\affiliation{Department of Applied Physics, Hokkaido University, Sapporo 060-8628, Japan}

\begin{abstract}
Discrete quantum walks are periodically driven systems with discrete time evolution. 
In contrast to ordinary Floquet systems, no microscopic Hamiltonian exists, and the one-period time evolution is given directly by a series of unitary operators.
Regarding each constituent unitary operator as a discrete time step, 
we formulate discrete space-time symmetry in quantum walks
and evaluate the corresponding symmetry protected topological phases.
In particular, we study chiral and/or time-glide symmetric topological quantum walks in this formalism.
Due to discrete nature of time-evolution,
the topological classification is found to be different from that in conventional Floquet systems.
As a concrete example, we study a two-dimensional quantum walk having both chiral and time-glide symmetries, and identify the anomalous edge states protected by these symmetries.
\end{abstract}


\maketitle
\section{introduction}
\label{sec:introduction}

Recently, topological phases in periodically driven systems (Floquet systems) have attracted much attention \cite{rahav2003effective,oka2009photovoltaic,kitagawa2010topological,kitagawa2011transport,lindner2011floquet,
wang2013observation,gometzleon2013floquet,rechtsman2013photonic,rudner2013anomalous,asboth2014chiral,
perezpiskunow2014floquet,nathan2015topological,takasan2017laser,potirniche2017floquet,roy2017floquet,
morimoto2017floquet,nakata2019laser,oka2019floquet,higashikawa2019floquet}. 
A large number of approaches to realize Floquet topological phases have been theoretically proposed \cite{oka2009photovoltaic,kitagawa2011transport,lindner2011floquet,
perezpiskunow2014floquet,takasan2017laser,potirniche2017floquet,nakata2019laser}
and experiments have been carried out \cite{wang2013observation,rechtsman2013photonic} to engineer Floquet systems with non-trivial topological phases. 
In many cases, topological phases in Floquet systems are understood in analogy with topological phases in static systems. Actually, the topological classification of Floquet gapped phases with non-spatial AZ symmetry is essentially the same as the static counterpart \cite{nathan2015topological,roy2017floquet,schnyder2008classification}. 
On the other hand, there have been also research interests in unique topological phases intrinsic to Floquet systems, in both gapped and gapless phases  \cite{rudner2013anomalous,kitagawa2010topological,asboth2014chiral,morimoto2017floquet,
higashikawa2019floquet}. 
For instance, anomalous gapless states are possible in both edge and bulk due to the periodicity in the quasi-energy spectrum \cite{rudner2013anomalous,higashikawa2019floquet}. 
Furthermore, recently, it was pointed out that there exist non-trivial topological phases protected by space-time symmetry specific to time-dependent systems such as time-glide symmetry \cite{morimoto2017floquet, xu2018space}, although the experimental realization of such systems has not been done yet. 

Among a lot of theoretical proposals and experimental settings for exploring Floquet topological phases, discrete quantum walks provide a versatile platform for this purpose \cite{rudner2009topological,
kitagawa2010exploring,
obuse2011topological,
kitagawa2012observation,
asboth2012symmetries,
asboth2013bulk,
tarasinski2014scattering,
asboth2015edge,
obuse2015unveiling,
cardano2016statistical,
gerasimenko2016attractor,
cedzich2016bulk,
barkhofen2017measuring,
verga2017edge,
endo2017sensitivity,
zhan2017detecting,
xiao2017observation,
barkhofen2018supersymmetric,
cedzich2018topological,
cedzich2018complete,
chen2018observation,
xiao2018higher,
nitsche2019eigenvalue,
mochizuki2020stability}.
The quantum walks are realizable in various physical settings, where the topological numbers can be easily tuned and the exact description of time evolution is easily obtained. 
Owing to these features, various topological edge states have been observed in quantum walks \cite{kitagawa2012observation,xiao2017observation,chen2018observation,xiao2018higher},
which include ones unique to Floquet systems, such as $\pi$ energy edge states in one-dimensional quantum walks \cite{barkhofen2017measuring,xiao2017observation,xiao2018higher} and anomalous edge states with energy winding in two-dimensional (2D) quantum walks \cite{chen2018observation}. 
While discrete quantum walks have experimental feasibility as mentioned above, their space-time symmetry and the corresponding symmetry protected topological phases have not been discussed yet.

It should be noted here that there exist essential distinctions between oridinary Floqeut systems and discrete quanutm walks, in spite of their similarity:
In conventional Floquet systems, the time-evolution operator $U({\bm k},t_1 \rightarrow t_2)$ is given by the microscopic Hamiltonian $H({\bm k}, t)$,
\begin{align}
U({\bm k},t_1 \rightarrow t_2)
={\cal T}\exp\left[-i\int_{t_1}^{t_2} d\tau H({\bm k},\tau)
\right],    
\label{eq:FloquetU}
\end{align}
where ${\cal T}$ denotes time-ordering, and the one-period time-evolution is evaluated by $U({\bm k},0 \rightarrow T)$ with the time-period $T$ in the Hamiltonian, $H({\bm k}, t+T)=H({\bm k}, t)$. In this case, space-time symmetry is naturally defined as a symmetry operation on the space-time $({\bm k},t)$ of the Hamiltonian. 
In contrast, in discrete quantum walks, no microscopic Hamiltonian exists and the one-period time-evolution $U({\bm k})$ is given directly by a series of unitary operators $U_i$,
\begin{align}
U({\bm k})=\prod_i U_i({\bm k}).  
\label{eq:QuantumWalkU}
\end{align}
Therefore, space-time symmetry should be defined as an operation on $U_i({\bm k})$.
We also note that $U({\bm k})$ in quantum walks can be topologically different from $U({\bm k},0 \rightarrow T)$ in Floquet systems.
For instance, whereas
any $U({\bm k},0 \rightarrow t)$ in Eq. (\ref{eq:FloquetU}) should be smoothly deformed into $U({\bm k},0 \rightarrow 0)=1$ in terms of the winding number of $U({\bm k},0 \rightarrow t)$, no such topological constraint is assumed for $U_i({\bm k})$ and $U({\bm k})$ in Eq. (\ref{eq:QuantumWalkU}). 

In this paper, we formulate space-time symmetry in discrete quantum walks and explore the corresponding symmetry protected topological phases. 
Regarding $U_i({\bm k})$ as a discrete time step, we introduce a discrete version of 
space-time symmetry.
In particular, we study chiral and time-glide symmetries in this formalism.
Following the arguments in Ref. \cite{asboth2013bulk}, we identify chiral symmetry as time-reflection symmetry, which enables us to treat these symmetries in a unified manner.
By explicit construction of topological numbers, we also present topological classification of quantum walks in the presence of either or both symmetries. 
We find that quantum walks may host topological numbers other than those in conventional Floquet systems because of the aforementioned topological difference between $U({\bm k})$ and $U({\bm k},t_1 \rightarrow t_2)$.

We also study a model of 2D quantum walks which has both time-glide and chiral symmetries.
This model is not fully gapped, and thus it might not ideal to study a strong aspect of the topological phase.
Nevertheless, the system hosts a variety of edge states protected by these symmetries.
We specify relevant topological numbers for each edge state.
Through the simulation for the time evolution of probability of quantum walks,
we also find that anomalous edge states protected by time-glide chiral symmetry are well-localized on the boundary, even in the presence of bulk gapless modes.


This paper is organized as follows. In Sec. \ref{sec:symmetry}, we formulate discrete space-time symmetry in quantum walks. In particular, we consider chiral symmetry and time-glide symmetry of quantum walks and explain how to construct models with these symmetries. In Sec. \ref{sec:anormalous_topological_phases}, we argue anomalous topological phases in quantum walks, which are protected by chiral and/or time-glide symmetries. 
By constructing relevant topological numbers explicitly, we present the topological tables.
In Sec. \ref{subsec:model}, we examine a 2D model of quantum walks with time-glide and chiral symmetries, and show that there appear topologically protected edge states of which origin is these symmetries. We summarize our results in Sec. \ref{sec:summary}.

\section{Discrete time-glide Symmetry: Asymmetric Unit Construction}
\label{sec:symmetry}
\subsection{chiral symmetry = time-mirror symmetry}
\label{sec:chiral}
For preparation, we first consider chiral symmetry for quantum walks.
As is shown immediately, chiral symmetry reduces to a kind of space-time symmetry, which is a unique feature for quantum walks.  

In quantum walks, the time-evolution is given by a unitary time-step (Floquet) operator $U$,
\begin{align}
|\Psi(t+1)\rangle=U|\Psi(t)\rangle.
\label{eq:time-evolution}
\end{align}
The discrete time-evolution naturally introduces a periodic structure in time-direction, where the unit cell is given by $t\in [0,1]$. In this picture, $U$ can be identified as an operator acting on the unit cell.

Chiral symmetry for quantum walks is defined as
\begin{align}
\Gamma U \Gamma^{\dagger}=U^{\dagger},    
\label{eq:chiral}
\end{align}
where $\Gamma$ is a unitary operator with $\Gamma^2=1$.
For unitary time-evolution, it is recast into
\begin{align}
\Gamma U \Gamma^{\dagger}=U^{-1},
\label{eq:chiral_unitary}
\end{align}
which implies that the chiral operator $\Gamma$ flips a time-step of the evolution such as
\begin{align}
t\to -t.    
\label{eq:time-mirror}
\end{align}
Therefore, chiral symmetry for quantum walks can be regarded 
as time-mirror symmetry, which is an analogue of mirror reflection symmetry in time direction. 
Note that time-mirror symmetry has nothing to do with  ordinary time-reversal symmetry since $\Gamma$ is unitary.

A viewpoint of time-mirror symmetry enables us to provide a systematic construction of chiral symmetric models.
The key idea is to use the concept of asymmetric unit in crystallography. 
For a unit cell $t\in [0,1]$ in time, we introduce an asymmetric unit $t\in [0,1/2]$ which generates the whole unit cell by time-mirror reflection up to discrete time translation.  
Then, we assign a time-evolution operator $U_1$ with the asymmetric unit.
Since the asymmetric unit recovers the whole unit cell by time-mirror reflection,
$U_1$ also generates $U$.
Actually, applying time-mirror operation to the asymmetric unit, we obtain another time-unit $t\in [-1/2,0]\approx [1/2,1]$ called orbital, on which we generate 
a time-evaluation operator $U_2$ by
\begin{align}
\Gamma U_1\Gamma^{\dagger}=U_2^{\dagger}.
\label{eq:chiral_U12}
\end{align}
Gluing together $U_1$ and $U_2$, we have the full time-evaluation operator $U$ as
\begin{align}
U=U_2U_1.
\label{eq:full_time-evolution_operator}
\end{align}
The obtained $U$ automatically hosts chiral symmetry in Eq. (\ref{eq:chiral}).


\subsection{time-glide symmetry}
\label{sec:time-glide}
Now, we formulate time-glide symmetry for quantum walks.
Time-glide symmetry is an analog of glide symmetry with partial time translation replacing partial space translation. More precisely, time-glide operation is defined by
\begin{align}
&{\bm x}\to g{\bm x},
\quad t\to t+1/2
\label{eq:operation_spacetime_time-glide}
\end{align}
on space-time, where $g{\bm x}=(x_1,\dots, x_{d-1}, -x_{d})$.
On a microscopic Hamiltonian $H({\bm k},t)=H({\bm k},t+1)$ of a conventional Floquet system, time-glide symmetry is defined as ${\cal G}_{\rm T}H({\bm k},t){\cal G}_{\rm T}^\dagger=H(g{\bm k},t+1/2)$ \cite{morimoto2017floquet} where ${\cal G}_{\rm T}$ is a unitary operator with $\mathcal{G}_{\rm T}^2=1$ and $g{\bm k}=(k_1, \dots, k_{d-1}, -k_d)$. In terms of the time-evolution operator in Eq. (\ref{eq:FloquetU}), it reads ${\cal G}_{\rm T}U({\bm k},0\rightarrow1/2) {\cal G}_{\rm T}^{\dagger}=U(g{\bm k},1/2\rightarrow1)$. On the other hand, a discrete quantum walk does not have a microscopic Hamiltonian. Thus, instead, to define time-glide symmetry, we use the idea of asymmetric unit, again.
For a time-unit cell $t\in [0,1]$, we have the asymmetric unit $t\in [0,1/2]$ 
and its orbital $t\in [1/2,1]$ for time-glide symmetry.
%
Then, we assign time-evolution operators $U_1$ and $U_2$ with the asymmetric unit and its orbital, respectively, where the half-time step in Eq. (\ref{eq:operation_spacetime_time-glide}) is interpreted as the interchange of $U_1$ and $U_2$. 
Time-glide symmetry is defined as
\begin{align}
{\cal G}_{\rm T} U_1({\bm k}){\cal G}_{\rm T}^{\dagger}=U_2(g{\bm k}),
\quad U=U_2U_1,
\label{eq:time-glide}
\end{align}
with unitary and Hermitian ${\cal G}_{\rm T}$, where $U_i({\bm k})$ is the momentum-space representation of $U_i$ and $U$ is the full time-evolution operator.

\subsection{time-glide and chiral symmetries}
\label{sec:time-glide+chiral}

Finally, we consider quantum walks that have both chiral symmetry (\ref{eq:chiral}) 
and time-glide symmetry (\ref{eq:time-glide}).
The time-glide operator ${\cal G}_{\rm T}$ and the chiral operator $\Gamma$ satisfy commutation or anticommutation relation, $\Gamma{\cal G}_{\rm T}=\eta_C{\cal G}_{\rm T}\Gamma$ with $\eta_C=\pm 1$.

The asymmetric unit is useful again to construct models with these symmetries.
The symmetry operations act on space-time as
\begin{align}
{\bm x}\to g{\bm x}, \quad t\to t+1/2, 
\end{align}
for time-glide symmetry, and 
\begin{align}
t\to -t, 
\end{align}
for chiral symmetry. The asymmetric unit and its orbitals under these symmetries are given by $t\in[0,1/4]$, $t\in [1/4,1/2]$, $t\in [1/2,3/4]$, and $t\in [3/4,1]$.
Corresponding to them, 
we introduce time-evolution operators $V_1$, $V_2$, $V_3$, and $V_4$, which are related to each other by symmetry
\begin{align}
\Gamma V_1 \Gamma^{\dagger}=V_4^{\dagger}, 
\quad \Gamma V_2 \Gamma^{\dagger}=V_3^{\dagger},    
\label{eq:chiral_time-glide_1}
\end{align}
and
\begin{align}
{\cal G}_{\rm T} V_1({\bm k}){\cal G}_{\rm T}^{\dagger}=V_3(g{\bm k}),\quad
{\cal G}_{\rm T} V_2({\bm k}){\cal G}_{\rm T}^{\dagger}=V_4(g{\bm k}).
\label{eq:chiral_time-glide_2}
\end{align}
Note that $V_1$ for the asymmetric unit can generate the time-evolution operators for the orbitals, $V_2$, $V_3$, and $V_4$.
For instance, applying the time-glide operator and the chiral one to $V_1$, we have 
\begin{align}
V_2({\bm k})=\Gamma{\cal G}_{\rm T} V_1^{\dagger}(g{\bm k}) {\cal G}_{\rm T}^{\dagger}\Gamma^{\dagger}.
\end{align}
The full time-evolution operator is given by 
\begin{align}
U=V_4V_3V_2V_1,  
\label{eq:V4V3V2V1}
\end{align}
which consists of the first half period part $U_1$ and the second one $U_2$,
\begin{align}
U=U_2 U_1,
\quad
U_1=V_2 V_1, \quad U_2=V_4 V_3. 
\label{eq:U2U1_V2V1_V4V3}
\end{align}

One can easily check that $U$ satisfies both chiral symmetry (\ref{eq:chiral}) and time-glide one (\ref{eq:time-glide}).
In particular, we can reproduce the relations (\ref{eq:chiral_U12}) and (\ref{eq:time-glide}).

\section{Anomalous Topological Phases}
\label{sec:anormalous_topological_phases}

The (bulk) quasi-spectrum $\varepsilon_m({\bm k})$ of a quantum walk is given by
\begin{align}
U({\bm k})|u_m({\bm k})\rangle=e^{-i\varepsilon_m({\bm k})}|u_m({\bm k})\rangle,
\label{eq:eigenvalue-equation}
\end{align}
where $m$ is the band index and $\ket{u_m({\bm k})}$ is the corresponding eigenstate.
When the quasi-spectrum has a gap at $\varepsilon=\varepsilon_{\rm b}$ where the $\varepsilon$ is a specific quasi-energy, we can define the single-valued Floquet Hamiltonian $H_{\rm F}({\bm k})$ through
\begin{align}
H_{\rm F}({\bm k})=i\ln U({\bm k}),
\label{eq:Floquet-Hamiltonian}
\end{align}
where the branch cut of the logarithm is placed in the gap.
Then, if there is another gap of the quasi-spectrum at $\varepsilon=\varepsilon_{\rm g}$ which is apart from the branch cut $\varepsilon_{\rm b}$ with a region in which the quasi-spectrum exists between $\varepsilon_{\rm b}$ and $\varepsilon_{\rm g}$, $H_{\rm F}({\bm k})$ also has a gap in a usual sense; the spectrum of $H_{\rm F}({\bm k})$ is separated above and below the gap at $\varepsilon_{\rm g}$. 
 Under this situation, the system can be a topological insulator in a manner similar to an ordinary static case and thus we may have topologically protected gapless boundary states in the gap at $\varepsilon=\varepsilon_{\rm g}$.
On the other hand, even when $H_{\rm F}({\bm k})$ itself does not have a gap at any $\varepsilon_{\rm g}(\neq\varepsilon_{\rm b}$), 
there could be gapless boundary states 
at the branch cut $\varepsilon=\varepsilon_{\rm b}$.
The latter boundary states (so called anomalous Floquet edge states) are intrinsic to dynamical systems, 
which can not be explained by the topology of $H_{\rm F}({\bm k})$.
Below, using the asymmetric unit construction of quantum walks, we introduce topological numbers relevant to the anomalous gapless boundary states.
We summarize the results in Tables. \ref{table:chiral}, \ref{table:time-glide}, and \ref{table:time-glide_chiral}. One of chiral winding numbers for each gap in Table. \ref{table:chiral} is unique to discrete quantum walks, since two winding numbers reduce to one in ordinary Floquet systems  described by microscopic Hamiltonians. This is because we can always smoothly deform $U({\bm k},t)$ in Eq. (\ref{eq:FloquetU}) to $U({\bm k},0)=1$, while $U({\bm k})$ in Eq. (\ref{eq:QuantumWalkU}) has no such constraint. In the same reason, in Table. \ref{table:time-glide_chiral}, one of chiral winding numbers and one of time-glide Chern numbers for $(\eta_C,\varepsilon)=(+1,\pi)$ and $(-1,0)$ are peculiar to discrete quantum walks.

\subsection{chiral symmetric case}
\label{sec:chiralcase}

We first derive a general property of $U=U_2U_1$ with $\Gamma U_1 \Gamma^{\dagger}=U_2^{\dagger}$.
Without loss of generality, we here take the basis where $\Gamma$ is diagonal,
\begin{align}
\Gamma=\left(
\begin{array}{cc}
1& 0 \\
0     & -1
\end{array}
\right)    
\end{align}
with the $n\times n$ identity matrix $1$.
If we write $U_1$ in this basis as
\begin{align}
U_1=\left(
\begin{array}{cc}
\alpha & \beta \\
\gamma & \delta
\end{array}
\right),
\label{eq:U1}
\end{align}
where $\alpha$, $\beta$, $\gamma$, $\delta$ are $n\times n$ matrix functions of ${\bm k}$ (here the ${\bm k}$-dependence is implicit), then
the chiral symmetry yields
\begin{align}
U_2=\left(
\begin{array}{cc}
\alpha^{\dagger} & -\gamma^{\dagger} \\
-\beta^{\dagger}     & \delta^{\dagger}
\end{array}
\right),
\label{eq:U2_chiral}
\end{align}
and thus $U$ is given by
\begin{align}
U=\left(
\begin{array}{cc}
\alpha^{\dagger}\alpha-\gamma^{\dagger}\gamma     & \alpha^{\dagger}\beta-\gamma^{\dagger}\delta \\
-\beta^{\dagger}\alpha+\delta^{\dagger}\gamma    & -\beta^{\dagger}\beta+\delta^{\dagger}\delta
\end{array}
\right).
\label{eq:U_chiral}
\end{align}
From $U_1^{\dagger}U_1=1$, we also have 
\begin{align}
\alpha^{\dagger}\alpha+\gamma^{\dagger}\gamma=1, \quad \alpha^{\dagger}\beta+\gamma^{\dagger}\delta=0, 
\nonumber\\
\beta^{\dagger}\alpha+\delta^{\dagger}\gamma=0, \quad \beta^{\dagger}\beta+\delta^{\dagger}\delta=1,
\label{eq:unitaryU1}
\end{align}
which lead to
\begin{align}
U-1=\left(
\begin{array}{cc}
-2\gamma^{\dagger}\gamma     & 2\alpha^{\dagger}\beta \\
-2\beta^{\dagger}\alpha     & -2\beta^{\dagger}\beta
\end{array}
\right)
\label{eq:U-1}
\end{align}
and
\begin{align}
U+1=\left(
\begin{array}{cc}
2\alpha^{\dagger}\alpha     & 2\alpha^{\dagger}\beta \\
-2\beta^{\dagger}\alpha     & 2\delta^{\dagger}\delta
\end{array}
\right).
\label{eq:U+1}
\end{align}

Equations (\ref{eq:U-1}) and (\ref{eq:U+1}) imply the following lemma.

\vspace{2ex}
\noindent
{\bf Lemma}
$U$ has a quasi-spectrum gap at $\varepsilon=0$ $(\varepsilon=\pi)$
if and only if ${\rm det} \gamma \cdot {\rm det}\beta\neq 0$ (${\rm det} \alpha \cdot{\rm det}\delta\neq 0$).

\vspace{2ex}
\noindent
Equivalently, Lemma can be rephrased as Lemma'.

\vspace{2ex}
\noindent
{\bf Lemma'}
$U$ has an eigenstate with $\varepsilon=0$ $(\varepsilon=\pi)$
if and only if ${\rm det} \gamma\cdot {\rm det}\beta=0$ (${\rm det}\alpha\cdot {\rm det}\delta=0$).
\vspace{2ex}

\noindent
Lemma' is proved as follows.
Consider an eigenstate $|0\rangle$ of $U$ with the quasi-energy $\varepsilon=0$,
\begin{align}
U|0\rangle=|0\rangle.        
\label{eq:U0}
\end{align}
Here $|0\rangle$ can be an eigenstate of $\Gamma$ at the same time:
Indeed, from chiral symmetry (\ref{eq:chiral}), we have $U|0\rangle=\Gamma^{\dagger} U^{\dagger} \Gamma |0\rangle$, and thus Eq. (\ref{eq:U0}) implies $U\Gamma |0\rangle=\Gamma|0\rangle$.
Therefore, by considering $(1\pm \Gamma)|0\rangle$ as $|0\rangle$, we have a simultaneous eigenstate of $U$ and $\Gamma$ with the eigenvalues $\varepsilon=0$ and $\Gamma=\pm 1$.

If $|0\rangle$ has the eigenvalue of $\Gamma=1$, $|0\rangle$ takes the form of $(\xi, 0)^T$.
Then, Eqs.(\ref{eq:U-1}) and (\ref{eq:U0}) lead to
\begin{align}
(U-1)|0\rangle=
\left( 
\begin{array}{c}
-2\gamma^{\dagger}\gamma  \\
-2\beta^{\dagger}\alpha     
\end{array}
\right)\xi=0. 
\label{eq:U-1_eigenequation}
\end{align}
Therefore, $\xi$ is non-trivial if and only if ${\rm det}\gamma^{\dagger}\gamma=0$ and ${\rm det}\beta^{\dagger}\alpha=0$. 
From Eq. (\ref{eq:unitaryU1}), the latter condition ${\rm det}\beta^{\dagger}\alpha=0$ follows from the former one ${\rm det}\gamma=0$, and thus we have $|0\rangle$ with $\Gamma=1$ if and only if ${\rm det}\gamma=0$.
In a similar manner, we can also prove that $|0\rangle$ with $\Gamma=-1$ exists if and only if ${\rm det}\beta=0$.
Combining these results, we find that $|0\rangle$ exists if and only if ${\rm det}\gamma\cdot{\rm det}\beta=0$.
Using Eq. (\ref{eq:U+1}), we can also prove that  an eigenstate of $U$ with $\varepsilon=\pi$ exists if and only if ${\rm det}\alpha\cdot {\rm det}\delta=0$.

The above Lemma implies that if the system hosts a gap at $\varepsilon=0$ ($\varepsilon=\pi$), $\gamma$ and $\beta$ ($\alpha$ and $\delta$) are elements of $GL(n, \mathbb{C})$.
Because of the non-trivial homotopy for $GL(n,\mathbb{C})$,
\begin{align}
\pi_d(GL(n,\mathbb{C}))
=\left\{
\begin{array}{cl}
\mathbb{Z},  &  \mbox{for odd $d$}\\
0, & \mbox{for even $d$}
\end{array}
\right.,
\label{eq:homotopy}
\end{align}
we can define the winding numbers \cite{ryu2010topological}
\begin{align}
    \nu_{d=2p+1}[u]=
    \frac{p!}{(2 \pi i)^{p+1}(2p+1)!}
    \int_{\mathrm{BZ}}
    \text{tr}\left[\left(u^{-1} d u\right)^{2p+1}\right],
\end{align}
where $u=\alpha,\beta,\gamma$ and $\delta$ for $d=2p+1$ with integer $p$. 
The  winding numbers $\nu_d[\gamma]$ and $\nu_d[\beta]$ ($\nu_d[\alpha]$ and $\nu_d[\delta]$) for $\gamma$ and $\beta$ ($\alpha$ and $\delta$)
are defined if the system has a gap at $\varepsilon=0$ ($\varepsilon=\pi$) and the space dimension $d$ is odd.
From the bulk-boundary correspondence,  we expect the existence of gapless boundary states in the gap at $\varepsilon=0$ ($\varepsilon=\pi$) if $\nu_d[\gamma]$ and/or $\nu_d[\beta]$ ($\nu_d[\alpha]$ and/or $\nu_d[\delta]$) are non-zero. The bulk-boundary correspondence is proved when some restrictions on winding numbers are imposed \cite{asboth2013bulk,asboth2014chiral}.
It should be noted here that only a single gap either at $\varepsilon=0$ or $\varepsilon=\pi$
is required to define these winding numbers, thus the obtained gapless boundary states are intrinsic to dynamical systems.

\begin{table}[tb]
\caption{Topological numbers for anomalous topological phases of chiral symmetric quantum walks. The superscript and the argument of $\mathbb{Z}$ specify the topological number:
$\mathbb{Z}^{({\rm W})}[\alpha]$ indicates the chiral winding number $\nu_{d}(\alpha)$. 
}
\begin{tabular}{ccccccc}\hline \hline
\quad&Gap&\quad\quad &odd $d$ &\quad\quad & even $d$ &\quad \\ \hline
&$\varepsilon=0$&&$\mathbb{Z}^{(\rm W)}[\beta]\oplus\mathbb{Z}^{(\rm W)}[\gamma]$&&0 &\\
&$\varepsilon=\pi$& &$\mathbb{Z}^{(\rm W)}[\alpha]\oplus\mathbb{Z}^{(\rm W)}[\delta]$&& 0& 
\\\hline \hline
\end{tabular}
\label{table:chiral}
\end{table}

\subsection{time-glide symmetric case}
\label{sec:timeglidecase}

For a time-glide symmetric system, we can define a topological number if $U({\bm k})$ has a gap of the quasi-energy spectrum on the glide symmetric plane.  
To see this, consider $U({\bm k})$ on the glide symmetric plane at ${\bm k}={\bm k}_{\rm G}$, 
where ${\bm k}_{\rm G}=(k_1,\dots, k_{d-1}, 0)$ or $(k_1, \dots, k_{d-1}, \pi)$. 
From Eq. (\ref{eq:time-glide}), we have 
\begin{align}
U({\bm k}_{\rm G})={\cal G}_{\rm T}U_1({\bm k}_{\rm G}){\cal G}_{\rm T}U_1({\bm k}_{\rm G}),   
\label{eq:UkG}
\end{align}
where we used $\mathcal{G}_{\rm T}^\dagger=\mathcal{G}_{\rm T}$. This relation leads to
\begin{align}
H_{\rm F}({\bm k}_{\rm G})=2H_1({\bm k}_{\rm G}),
\label{eq:HFH1}
\quad \mbox{(mod 2$\pi$)}, 
\end{align}
where 
\begin{align}
&H_{\rm F}({\bm k}_{\rm G})=i\ln{U({\bm k}_{\rm G})},
\label{eq:lnHF}\\
&H_1({\bm k}_{\rm G})=i\ln{{\cal G}_{\rm T}U_1({\bm k}_{\rm G})}.
\label{eq:lnH1}
\end{align}
When $U({\bm k}_{\rm G})$ has a gap at $\varepsilon=\varepsilon_{\rm b}$, 
from Eq. (\ref{eq:HFH1}), we can regard $H_1({\bm k}_{\rm G})$ as a single valued matrix function with the branch cut of the logarithm in Eq. (\ref{eq:lnH1}) at $\varepsilon=\varepsilon_{\rm b}/2$, which has a gap at $\varepsilon=\varepsilon_{\rm b}/2+\pi$. Therefore, $H_1({\bm k}_{\rm G})$ defines an insulator and the corresponding topological number. 

Since no symmetry is imposed on $H_1({\bm k}_{\rm G})$, the relevant topological number is the Chern number \cite{nakahara2003geometry} on ${\bm k}_{\rm G}$, which we call time-glide Chern number:
\begin{align}
    Ch^{\text{TG}}_{p}=\frac{1}{p!}\left(\frac{i}{2 \pi}\right)^{p} \int_{\mathrm{BZ}|k_d=0/\pi} \operatorname{tr} \mathcal{F}^{p},
\label{eq:tgCh}
\end{align}
where $d=2p+1$ and $\mathcal{F}$ is the curvature
\begin{align}
    \mathcal{F}=d\mathcal{A}+\mathcal{A}^2, \quad [\mathcal{A}]_{lm}=\braket{\psi_l|d \psi_m},
\end{align}
for eigenstates $\ket{\psi_l}, \ket{\psi_m}$ of $H_1({\bm k}_{\rm G})$ with eigenenergies $\varepsilon_l,\varepsilon_m \in [\varepsilon_{\rm b}/2,\varepsilon_{\rm b}/2+\pi]$.
The trace in Eq. (\ref{eq:tgCh}) is taken for all the eigenstates within $ [\varepsilon_{\rm b}/2, \varepsilon_{\rm b}/2+\pi]$.

The time-glide Chern number is well-defined if the dimension of the glide symmetric plane is even (namely if $d$ is odd).
The time-glide Chern number cannot change as long as $H_{\rm F}({\bm k}_{\rm G})$ keeps the branch cut at $\varepsilon=\varepsilon_{\rm b}$ open. 
Therefore, if it is non-zero and the system has a boundary keeping the time-glide symmetry, gapless boundary states appear at $\varepsilon=\varepsilon_{\rm b}$.
On the other hand, if $d$ is even, there is no topological number for $H_1({\bm k}_{\rm G})$.

Here we should note that $H_{\rm F}({\bm k}_{\rm G})$ can be gapless even if $H_1({\bm k}_{\rm G})$ has a gap. 
Then, the ordinary Chern number of $H_{\rm F}({\bm k}_{\rm G})$ can change its value keeping the branch cut at $\varepsilon=\varepsilon_{\rm b}$ open. Therefore, the Chern number of $H_{\rm F}({\bm k}_{\rm G})$ is not relevant to anomalous topological numbers which cannot be changed as long as the branch cut at $\varepsilon=\varepsilon_{\rm b}$ is open.

\begin{table}[b]
\caption{Topological numbers for anomalous topological phases of time-glide symmetric quantum walks. The superscript and the argument of $\mathbb{Z}$ specify the topological number. $\mathbb{Z}^{\rm TGCh}[H_1]$ indicates the time-glide Chern number of $H_1$. While time-glide Chern numbers are defined at $k_d=0$ and $\pi$, we show only strong index corresponding to the difference of them, where Table. \ref{table:time-glide_chiral} is made in the same principle.}
\begin{tabular}{ccccccc}\hline \hline
\quad&Gap&\quad\quad &odd $d$ &\quad\quad & even $d$ &\quad \\ \hline
&$\varepsilon=\varepsilon_{\rm b}$&&$\mathbb{Z}^{(\rm TGCh)}[H_1]$&& 0&
\\\hline \hline
\end{tabular}
\label{table:time-glide}
\end{table}


\subsection{time-glide and chiral symmetric case}

Only gaps at $\varepsilon=0, \pi$ are consistent with the coexistence of time-glide and chiral symmetries.
In the presence of either of these gaps, 
topological numbers can be constructed from the half period part of the time-evolution operator.
For $U=V_4V_3V_2V_1$, 
the first (second) half period part is $U_1=V_2V_1$ ($U_2=V_4V_3$). 
From Eqs. (\ref{eq:chiral_time-glide_1}) and (\ref{eq:chiral_time-glide_2}), 
they satisfy  
\begin{align}
\Gamma U_{1}({\bm k})\Gamma^{\dagger}=U_{2}^{\dagger}({\bm k}),
\quad
{\cal G}_{\rm T}U_{1}({\bm k}){\cal G}_{\rm T}^\dagger=U_{2}(g{\bm k}).
\label{eq:U1U2}
\end{align}

When the space dimension $d$ is odd, 
using the first (second) equation in Eq. (\ref{eq:U1U2}), $U_1$ gives the winding numbers (the time-glide Chern number) as shown in Sec.\ref{sec:chiralcase} (Sec.\ref{sec:timeglidecase}). 
However, the second (first) equation in Eq. (\ref{eq:U1U2}) gives additional constraints on the winding numbers (the time-glide Chern number).
The constraints depend on the commutation relation between $\Gamma$ and ${\cal G}_{\rm T}$. We discuss the constraints in Secs. \ref{sec:tgc_odd_commute} and \ref{sec:tgc_odd_anticommute}.
On the other hand, when $d$ is even,  $U_1$ does not provide any strong topological number. 

We can also consider another time-evolution operator 
\begin{align}
U'=U_2'U_1',
\label{eq:U_prime}
\end{align}
where  $U_1'$ and $U_2'$ are defined as
\begin{align}
U_1'=V_1V_4,\ \ U_2'=V_3V_2,
\label{eq:U12_prime}
\end{align}
respectively. Since $U'$ is unitary equivalent to $U$, {\it i.e.} $U'=V_4^{\dagger}U V_4$,
$U'$ has the same quasi-spectra as $U$. 
As a result, $U'$ also provides topological numbers for anomalous edge states. 
From Eqs. (\ref{eq:chiral_time-glide_1}) and (\ref{eq:chiral_time-glide_2}), 
the first and the second half period parts obey
\begin{align}
\Gamma U'_{2}({\bm k})\Gamma^{\dagger}=U'^{\dagger}_{2}({\bm k}), \quad
&\Gamma U'_{1}({\bm k})\Gamma^{\dagger}=U'^{\dagger}_{1}({\bm k}),
\label{eq:U'symmetry0}\\
{\cal G}_{\rm T}U'_{2}({\bm k}){\cal G}_{\rm T}^\dagger&=U_1'(g{\bm k}).
\label{eq:U'symmetry}
\end{align}
Then, we also obtain
\begin{align}
\Gamma {\cal G}_{\rm T}U_1'({\bm k}){\cal G}_{\rm T}^\dagger\Gamma^\dagger=U_2'^{\dagger}(g{\bm k}),   
\label{eq:time-glide_chiral_0}
\end{align}
which implies a variation of chiral symmetry (we call it time-glide chiral symmetry) on the glide symmetric plane ${\bm k}={\bm k}_{\rm G}$,
\begin{align}
\Gamma {\cal G}_{\rm T}U_1'({\bm k}_{\rm G}){\cal G}_{\rm T}^\dagger\Gamma^\dagger=U_2'^{\dagger}({\bm k}_{\rm G}).
\label{eq:time-glide_chiral}
\end{align}

When $d$ is odd, from Eq. (\ref{eq:U'symmetry}), $U_1'$ gives the time-glide Chern number in a manner similar to $U_1$.
On the other hand, when $d$ is even, $U'_1({\bm k}_{\rm G})$ defines a $(d-1)$-dimensional winding number on the $(d-1)$-dimensional glide symmetric plane ${\bm k}={\bm k}_{\rm G}$ by using the time-glide chiral symmetry in Eq. (\ref{eq:time-glide_chiral}).
As we shall discuss in Secs.\ref{sec:tgc_odd_commute}, \ref{sec:tgc_even_commute}, \ref{sec:tgc_odd_anticommute}, and \ref{sec:tgc_even_anticommute},
both the topological numbers are subject to the constraints originating from the remaining symmetry.

\subsubsection{Case with odd space dimension $d$ and $[\Gamma,{\cal G}_{\rm T}]=0$}
\label{sec:tgc_odd_commute}

We first consider the case with odd $d$ and $[\Gamma, {\cal G}_{\rm T}]=0$.
When $\Gamma$ and ${\cal G}_{\rm T}$ commute with each other, we can take the basis where both $\Gamma$ and ${\cal G}_{\rm T}$ are block-diagonal,  
\begin{align}
\Gamma=
\begin{pmatrix}
1 & 0\\
0 & -1
\end{pmatrix},
\quad
{\cal G}_{\rm T}=
\begin{pmatrix}
g_+ & 0\\
0 & g_-
\end{pmatrix}.
\label{eq:basis_odd_commute}
\end{align}
Denoting $U_1({\bm k})$ in the basis of Eq. (\ref{eq:basis_odd_commute}) as
\begin{align}
U_1=
\begin{pmatrix}
\alpha & \beta \\
\gamma & \delta\end{pmatrix},
\end{align}
we have the following relations from Eq. (\ref{eq:U1U2}),
\begin{align}
g_+ \alpha({\bm k}) g_+^{\dagger}=\alpha^{\dagger}(g{\bm k}), \quad
g_+ \beta({\bm k}) g_-^{\dagger}=-\gamma^{\dagger}(g{\bm k}),
\nonumber\\
g_- \gamma({\bm k}) g_+^{\dagger}=-\beta^{\dagger}(g{\bm k}), \quad
g_- \delta({\bm k}) g_-^{\dagger}=\delta^{\dagger}(g{\bm k}).
\label{eq:constraint_abcd_odd_commute}
\end{align}
The relations restrict possible winding numbers of the system:
When $d$ is odd and the system has a gap at $\varepsilon=0$ ($\varepsilon=\pi$), from Lemma in Sec. \ref{sec:chiral}, 
the winding numbers $\nu[\gamma]$ and $\nu[\beta]$ ($\nu[\alpha]$ and $\nu[\delta]$) are well-defined.
Equation (\ref{eq:constraint_abcd_odd_commute}) gives the constraint $\nu[\beta]=\nu[\gamma]$. (No constraint is obtained for $\nu[\alpha]$ and $\nu[\delta]$.) 

For odd $d$, the system also has the time-glide Chern number, namely the Chern number of $H_1({\bm k}_{\rm G})=i\ln{\cal G}_{\rm T}U_1({\bm k}_{\rm G})$.
However, there exists an additional constraint:
From Eq. (\ref{eq:U1U2}), we also have 
\begin{align}
\Gamma {\cal G}_{\rm T} U_1({\bm k}_{\rm G}){\cal G}_{\rm T}^{\dagger}\Gamma^\dagger=U_1^{\dagger}({\bm k}_{\rm G}).   
\label{eq:time-glide_chiral_odd_commute}
\end{align}
In the present case with $[\Gamma, {\cal G}_{\rm T}]=0$, Eq. (\ref{eq:time-glide_chiral_odd_commute}) implies the chiral symmetry for $H_1({\bm k}_{\rm G})$, 
\begin{align}
\Gamma e^{-iH_1({\bm k}_{\rm G})}\Gamma^{-1} =e^{iH_1({\bm k}_{\rm G})},
\label{eq:constraint_H1_odd_commute}
\end{align}
which indicates that the time-glide Chern number is identically zero when the system has a gap at $\varepsilon=0$. 
Actually, in this case, $H_1({\bm k})$ has two gaps at $\varepsilon=0$ and $\pi$, and thus the chiral symmetry in Eq. (\ref{eq:constraint_H1_odd_commute}) interchanges two bands of $H_1({\bm k}_{\rm G})$ separated by these gaps. As a result, the Chern number of $H_1({\bm k}_{\rm G})$ becomes zero. On the other hand, the time-glide Chern number may survive if the system has a gap only at $\varepsilon=\pi$. 
In this case, 
$H_1({\bm k}_{\rm G})$ has gaps at $\varepsilon=\pi/2$ and $3\pi/2$.
Since chiral symmetry in Eq. (\ref{eq:constraint_H1_odd_commute}) maps bands of $H_1({\bm k}_{\rm G})$ inside these gaps to themselves,
Eq. (\ref{eq:constraint_H1_odd_commute}) does not give any constraint on the time-glide Chern number. 

The system may have another time-glide Chern number:
From Eq. (\ref{eq:U'symmetry}) and $\mathcal{G}_{\rm T}^\dagger=\mathcal{G}_{\rm T}$, $U'({\bm k}_{\rm G})$ is recast into 
\begin{align}
U'({\bm k}_{\rm G})
={\cal G}_{\rm T}U_1'({\bm k}_{\rm G}){\cal G}_{\rm T}U_1'({\bm k}_{\rm G}).
\label{eq:U_prime_odd_commute}
\end{align}
Thus, in a similar manner to $U_1({\bm k}_{\rm G})$, we can define the time-glide Chern number as the Chern number of 
$H_1'({\bm k}_{\rm G})$ defined below,
\begin{align}
H_1'({\bm k}_{\rm G})=i\ln {\cal G}_{\rm T}U_1'({\bm k}_{\rm G}). 
\label{eq:H1prime}
\end{align}
Note that Eq. (\ref{eq:U'symmetry0}) and $[\Gamma, {\cal G}_{\rm T}]=0$ lead to a variation of chiral symmetry for $H'_1({\bm k}_{\rm G})$,
\begin{align}
\Gamma {\cal G}_{\rm T} e^{-iH_1'({\bm k}_{\rm G})} {\cal G}_{\rm T}^\dagger\Gamma^\dagger=e^{iH_1'({\bm k}_{\rm G})},
\label{eq:constraint_H1_prime_odd_commute}
\end{align}
which implies that the time-glide Chern number is zero when the system has a gap at $\varepsilon=0$.
Therefore, we have this second time-glide Chern number only when the system has a gap at $\varepsilon=\pi$. 

\subsubsection{Case with even space dimension $d$ and $[\Gamma, {\cal G}_{\rm T}]=0$}
\label{sec:tgc_even_commute}

When $d$ is even, using chiral symmetry in Eq. (\ref{eq:time-glide_chiral}) and Lemma in Sec.\ref{sec:chiralcase}, we can define a winding number, which we call time-glide chiral winding number,  from an element of $U_1'({\bm k}_{\rm G})$. 
This topological number is also subject to a constraint.
To see this, we take the basis where $\Gamma{\cal G}_{\rm T}$ is diagonal, 
\begin{align}
\Gamma {\cal G}_{\rm T}=
\begin{pmatrix}
1 & 0 \\
0 & -1
\end{pmatrix},
\label{eq:basis_even_commute}
\end{align}
and denote $U_1'({\bm k}_{\rm G})$ in this basis as
\begin{align}
U_1'=
\begin{pmatrix}
\alpha' & \beta' \\
\gamma' & \delta'
\end{pmatrix},
\label{eq:U1_prime_even_commute}
\end{align}
where the dependence of ${\bm k}_{\rm G}$ is implicit.
From time-glide chiral symmetry in Eq. (\ref{eq:time-glide_chiral}), $U_2'({\bm k}_{\rm G})$ is written as
\begin{align}
U_2'=
\begin{pmatrix}
\alpha'^{\dagger} & -\gamma'^{\dagger}\\
-\beta'^{\dagger} & \delta'^{\dagger}
\end{pmatrix}.  
\label{eq:U2_prime_even_commute}
\end{align}
Since ${\cal G}_{\rm T}$ commutes with $\Gamma {\cal G}_{\rm T}$, ${\cal G}_{\rm T}$ can be written as
\begin{align}
{\cal G}_{\rm T}=
\begin{pmatrix}
g'_+ & 0\\
0 & g_-'
\end{pmatrix},
\end{align}
and thus Eq. (\ref{eq:U'symmetry}) leads the constraints
\begin{align}
g'_+\alpha'g'^{\dagger}_+=\alpha'^{\dagger}, \quad g'_+\beta'g'^{\dagger}_-=-\gamma'^{\dagger},
\nonumber\\
g'_-\gamma'g'^{\dagger}_+=-\beta'^{\dagger}, \quad g_-'\delta'g'^{\dagger}_-=\delta'^{\dagger}.
\label{eq:constraint_abcd_even_commute}
\end{align}
Replacing $\Gamma$ and $U$ with $\Gamma{\cal G}_{\rm T}$ and $U'$, we can apply Lemma in Sec. \ref{sec:chiralcase} to the present case.
Thus, if $U'({\bm k}_{\rm G})$ has a gap at $\varepsilon=0$ ($\varepsilon=\pi$), we can define the $(d-1)$-dimensional winding numbers 
$\nu_{d-1}[\beta']$ and $\nu_{d-1}[\gamma']$ ($\nu_{d-1}[\alpha']$ and $\nu_{d-1}[\delta']$) in the $(d-1)$-dimensional glide symmetric subspace.
However, from Eq. (\ref{eq:constraint_abcd_even_commute}), it holds that $\nu_{d-1}[\alpha']=\nu_{d-1}[\delta']=0$ and $\nu_{d-1}[\beta']=-\nu_{d-1}[\gamma']$.
Consequently, we have a unique winding number $\nu_{d-1}[\beta']=-\nu_{d-1}[\gamma']$ if $U'({\bm k}_{\rm G})$ has a gap at $\varepsilon=0$.

\subsubsection{Case with odd space dimension $d$ and $\{\Gamma,{\cal G}_{\rm T}\}=0$}
\label{sec:tgc_odd_anticommute}

When $\Gamma$ and ${\cal G}_{\rm T}$ anti-commute with each other, we can take the basis where $\Gamma$ is diagonal and ${\cal G}_{\rm T}$ is off-diagonal,
\begin{align}
\Gamma=
\begin{pmatrix}
1 & 0\\
0 & -1
\end{pmatrix},
\quad
{\cal G}_{\rm T}=
\begin{pmatrix}
0 & g_+\\
g_- & 0
\end{pmatrix}.
\label{eq:basis_odd_anticommute}
\end{align}
Taking $U_1({\bm k})$ in this basis as
\begin{align}
U_1=
\begin{pmatrix}
\alpha & \beta \\
\gamma & \delta
\end{pmatrix},
\label{eq:U1_odd_anticommute}
\end{align}
we have the following constraints from Eq. (\ref{eq:U1U2}),
\begin{align}
g_+ \delta({\bm k}) g_+^{\dagger}=\alpha^{\dagger}(g{\bm k}), \quad
g_+ \gamma({\bm k}) g_-^{\dagger}=-\gamma^{\dagger}(g{\bm k}),
\nonumber\\
g_- \beta({\bm k}) g_+^{\dagger}=-\beta^{\dagger}(g{\bm k}), \quad
g_- \alpha({\bm k}) g_-^{\dagger}=\delta^{\dagger}(g{\bm k}).
\label{eq:constraint_abcd_odd_anticommute}
\end{align}
These relations restrict possible winding numbers of $U_1({\bm k})$: 
When $d$ is odd and the system has a gap at $\varepsilon=0$ ($\varepsilon=\pi$), from Lemma in Sec. \ref{sec:chiralcase}, 
the winding numbers $\nu_d[\beta]$ and $\nu_d[\gamma]$ ($\nu_d[\alpha]$ and $\nu_d[\delta]$) are well-defined.
Equation (\ref{eq:constraint_abcd_odd_anticommute}) gives no constraint for $\nu[\beta]$ and $\nu[\gamma]$, while it gives  $\nu[\alpha]=\nu[\delta]$.
For odd $d$, we can also define the time-glide Chern number, but it can be nonzero only when the system has a gap at $\varepsilon=0$.
Because of the anti-commutation relation $\{\Gamma, {\cal G}_{\rm T}\}=0$, Eq. (\ref{eq:time-glide_chiral_odd_commute}) leads to
\begin{align}
\Gamma e^{-i[H_1({\bm k}_{\rm G})-\pi/2]}\Gamma^{-1}=e^{i[H_1({\bm k}_{\rm G})-\pi/2]},
\label{eq:constraint_H1_odd_anticommute}
\end{align}
from which we find that the time-glide Chern number becomes zero when the system has a gap at $\varepsilon=\pi$.
We also have another time-glide Chern number, using $H_1'({\bm k}_{\rm G})$ in Eq. (\ref{eq:H1prime}).
In the present case, Eqs. (\ref{eq:U'symmetry0}) implies
\begin{align}
\Gamma {\cal G}_{\rm T} e^{-i[H_1'({\bm k}_{\rm G})-\pi/2]} {\cal G}_{\rm T}^\dagger\Gamma^\dagger=e^{i[H_1'({\bm k}_{\rm G})-\pi/2]},
\label{eq:constraint_H1_prime_odd_anticommute}
\end{align}
from which the time-glide Chern number becomes zero for a gap at $\varepsilon=\pi$.

\subsubsection{Case with even space dimension $d$ and $\{\Gamma, {\cal G}_{\rm T}\}=0$}
\label{sec:tgc_even_anticommute}

Finally, we discuss constraints for the time-glide chiral winding number, which is defined by using time-glide chiral symmetry in Eq. (\ref{eq:time-glide_chiral}).
Since $\Gamma{\cal G}_{\rm T}$ and ${\cal G}_{\rm T}$ anti-commute with each other, we can take the basis with
\begin{align}
\Gamma{\cal G}_{\rm T}=
\begin{pmatrix}
1 & 0\\
0 & -1
\end{pmatrix},
\quad
{\cal G}_{\rm T}=
\begin{pmatrix}
0 & g_+'\\
g_-' & 0
\end{pmatrix}.
\label{eq:basis_even_anticommute}
\end{align}
Denoting $U_1'({\bm k}_{\rm G})$ in this basis as
\begin{align}
U_1'=
\begin{pmatrix}
\alpha' & \beta'\\
\gamma' & \delta'
\end{pmatrix},
\label{eq:U1_prime_even_anticommute}
\end{align}
we have constraints
\begin{align}
g'_+\delta'g'^{\dagger}_+=\alpha'^{\dagger}, \quad g'_+\gamma'g'^{\dagger}_-=-\gamma'^{\dagger},
\nonumber\\
g'_-\beta'g'^{\dagger}_+=-\beta'^{\dagger}, \quad g_-'\alpha'g'^{\dagger}_-=\delta'^{\dagger},
\label{eq:constraint_even_anticommute}
\end{align}
and thus $\nu_{d-1}[\beta']=\nu_{d-1}[\gamma']=0$ and $\nu_{d-1}[\alpha']=-\nu_{d-1}[\delta']$. 
Noting that the presence of $\nu_{d-1}[\alpha']$ and $\nu_{d-1}[\delta']$ is ensured by the gap at $\varepsilon=\pi$,  
the time-glide chiral winding number is nonzero only when the system has a gap at $\varepsilon=\pi$.

\begin{table*}[bt]
\caption{Topological numbers for anomalous topological phases of time-glide and chiral symmetric quantum walks. The superscript and the argument of $\mathbb{Z}$ specify the corresponding topological number. $\eta_C=\pm 1$ in the left column represents whether $\Gamma$ and ${\cal G}_{\rm T}$ commute or anticummute with each other, $\Gamma{\cal G}_{\rm T}=\eta_C{\cal G}_{\rm T}\Gamma$.
$\mathbb{Z}^{({\rm W})}[\alpha]$, $\mathbb{Z}^{({\rm TGCh})}[H_1]$ and $\mathbb{Z}^{({\rm TGCW})}[\alpha']$ indicate the winding number $\nu_d[\alpha]$, the time-glide Chern number of $H_1$ and the time-glide chiral winding number $\nu_{d-1}[\alpha']$, respectively. 
}
\begin{tabular}{ccccccccc}\hline \hline
\quad &$\eta_{\rm C}$&\quad\quad&Gap&\quad\quad &odd $d$ &\quad\quad & even $d$ &\quad \\ \hline
&    +1&&$\varepsilon=0$&&$\mathbb{Z}^{(\rm W)}[\beta]$ && $\mathbb{Z}^{({\rm TGCW})}[\beta']$ \\
&&&$\varepsilon=\pi$& &$\mathbb{Z}^{(\rm W)}[\alpha]\oplus\mathbb{Z}^{(\rm W)}[\delta]\oplus\mathbb{Z}^{({\rm TGCh})}[H_1]\oplus\mathbb{Z}^{({\rm TGCh})}[H_1']$ && 0 &\\ \hline
&-1&&$\varepsilon=0$& &$\mathbb{Z}^{(\rm W)}[\beta]\oplus\mathbb{Z}^{(\rm W)}[\gamma]\oplus\mathbb{Z}^{({\rm TGCh})}[H_1]\oplus\mathbb{Z}^{({\rm TGCh})}[H_1']$ && 0& \\
&    &&$\varepsilon=\pi$&&$\mathbb{Z}^{(\rm W)}[\alpha]$ && $\mathbb{Z}^{({\rm TGCW})}[\alpha']$ &
\\\hline \hline
\end{tabular}
\label{table:time-glide_chiral}
\end{table*}

\section{2D model with time-glide and chiral symmetries}
\label{subsec:model}
As an example, we construct a concrete model with time-glide symmetry and chiral symmetry based on asymmetric unit construction in Sec. \ref{sec:symmetry}, and explore topological phases of the model. We examine quantum walks of particles with two internal states $\sigma=\pm$,
in the two-dimensional square lattice. The system is described by
\begin{align}
\ket{\Psi(t)}=\sum_{{\bm r},\sigma}
\psi_\sigma(\mbox{\boldmath $r$},t)\ket{\mbox{\boldmath $r$}}\otimes\ket{\sigma},
\label{eq:state}
\end{align}
where $\mbox{\boldmath $r$}=(x,y)$ represents the position of the walker in the lattice and $\sigma=\pm$ denotes their internal states. 
For the standard quantum walks, the time-evolution operator $U$ consists of the coin and shift operators. 
In the vector basis of $\ket{\sigma}$, $\ket{+}=(0,1)^\text{T}$ and $\ket{-}=(1,0)^\text{T}$, the coin operators $C_j(\theta)$ ($j=1,2$) are given by
\begin{align}
C_j(\theta)=\sum_{{\bm r}}\ket{{\bm r}}\bra{{\bm r}}e^{i\theta\sigma_j}=e^{i\theta\sigma_j}.
\label{eq:C12}
\end{align}
The coin operators rotate the internal states of walkers. The position of walkers is changed  by the shift operators $S_r^\pm$ ($r=x,y$),
\begin{align}
S_x^{\pm}=\sum_{\bm r}\left(
|{\bm r}\pm\hat{\bm x}\rangle\langle{\bm r}|
\otimes |\pm\rangle\langle \pm|
+|{\bm r}\rangle\langle{\bm r}|
\otimes |\mp\rangle\langle \mp|
\right),
\\
S_y^{\pm}=\sum_{\bm r}\left(
|{\bm r}\pm\hat{\bm y}\rangle\langle{\bm r}|
\otimes |\pm\rangle\langle \pm|
+|{\bm r}\rangle\langle{\bm r}|
\otimes |\mp\rangle\langle \mp|
\right),
\label{eq:Sxy}
\end{align}
where $\hat{\bm x}$ and $\hat{\bm y}$ are the unit vectors in the $x$ and $y$-directions, respectively.
In the momentum space, these operators are given by
\begin{align}
&C_j(\theta)=\bra{\bm k}C_j(\theta)\ket{\bm k}=e^{i\theta\sigma_j},
\label{eq:coin_k}\\
&S_r^{\pm}({\bm k})=\bra{\bm k}S_r^{\pm}\ket{\bm k}=e^{\pm i\frac{k_r}{2}}e^{i\frac{k_r}{2}\sigma_3}, 
\label{eq:shift_k}
\end{align}
where $r=x,y$.

Now we construct a two-dimensional quantum walk that supports both time-glide and chiral symmetries. As explained in Sec. \ref{sec:time-glide+chiral}, such a model is obtained systematically from a unitary operator $V_1$. We consider the following $V_1$,
\begin{align}
V_1=C_2(-\frac{\theta_2}{2})S_y^{-}C_1(-\frac{\theta_1}{2})
C_1(\frac{\pi}{4})S_x^{+}C_1(-\frac{\pi}{4})
\label{eq:V1_r}
\end{align}
with the time-glide and chiral operators given by
\begin{align}
{\cal G}_{\rm T}=\sigma_2, \quad \Gamma=\sigma_1.
\label{eq:gamma_gt}
\end{align}
In the momentum space, $V_1$ is represented as
\begin{align}
V_1({\bm k})&=C_2(-\frac{\theta_2}{2})S_y^{-}({\bm k})C_1(\frac{\pi}{4}-\frac{\theta_1}{2})
S_x^{+}({\bm k})C_1(-\frac{\pi}{4})   
\nonumber\\
&=e^{i\frac{k_x-k_y}{2}}C_2(-\frac{\theta_2}{2})Z(\frac{k_y}{2})C_1(-\frac{\theta_1}{2})Y(\frac{k_x}{2}),
\label{eq:V1_k}
\end{align}
where $Z(\frac{k_y}{2})=e^{i\frac{k_y}{2}\sigma_z}$ and $Y(\frac{k_x}{2})=e^{i\frac{k_x}{2}\sigma_y}$.
From Eqs. (\ref{eq:chiral_time-glide_1}) and (\ref{eq:chiral_time-glide_2}), we obtain
\begin{align}
V_2({\bm k})&=e^{-i\frac{k_x+k_y}{2}}Y(\frac{k_x}{2})C_1(-\frac{\theta_1}{2})Z(\frac{k_y}{2})C_2(-\frac{\theta_2}{2}),
\label{eq:V2}\\
V_3({\bm k})&=e^{i\frac{k_x+k_y}{2}}C_2(-\frac{\theta_2}{2})Z(\frac{k_y}{2})C_1(\frac{\theta_1}{2})Y(\frac{k_x}{2}),
\label{eq:V3}\\
V_4({\bm k})&=e^{-i\frac{k_x-k_y}{2}}Y(\frac{k_x}{2})C_1(\frac{\theta_1}{2})Z(\frac{k_y}{2})C_2(-\frac{\theta_2}{2}).
\label{eq:V4}
\end{align}
The full time-evolution operator $U$ and its unitary equivalent partner $U'$ are given by $U=V_4V_3V_2V_1$ and $U'=V_3V_2V_1V_4$, respectively.
In the present case, their first half period parts are 
\begin{align}
U_1({\bm k})&=e^{-ik_y}Y(\frac{k_x}{2})C_1(-\frac{\theta_1}{2})Z(\frac{k_y}{2})
\nonumber\\
&\times C_2(-\theta_2)Z(\frac{k_y}{2})C_1(-\frac{\theta_1}{2})Y(\frac{k_x}{2}),
\label{eq:U1_model}\\
U_1'({\bm k})&=C_2(-\frac{\theta_2}{2})Z(\frac{k_y}{2})C_1(-\frac{\theta_1}{2})
\nonumber\\
& \times Y(k_x)C_1(\frac{\theta_1}{2})Z(\frac{k_y}{2})C_2(-\frac{\theta_2}{2}).
\label{eq:U1_prime_model}
\end{align}

\begin{figure}[bt]
\begin{center}
\includegraphics[scale=0.82]{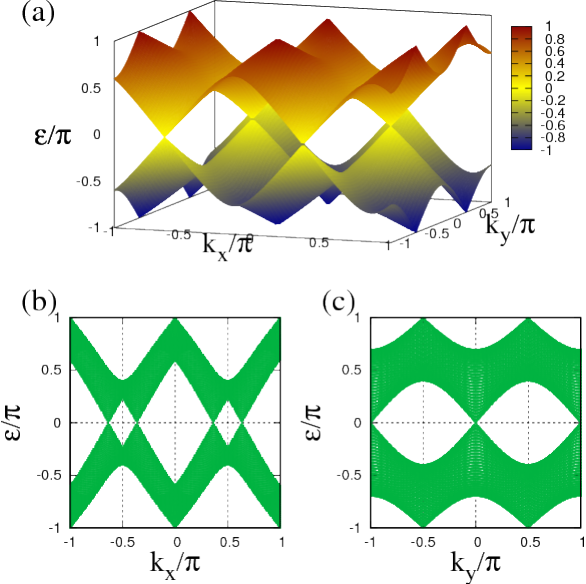}
\caption{(a) A dispersion relation
$\varepsilon_{\mbox{\boldmath $k$}}$ for the homogeneous system
with $\theta_1=-2\pi/7,\ \theta_2=-7\pi/10$. The same dispersion relation $\varepsilon_{\mbox{\boldmath $k$}}$ which is mapped on (b) $k_x$-$\varepsilon$ plane and (c) $k_y$-$\varepsilon$ plane. }
\label{fig:dispersion_homogeneous}
\end{center}
\end{figure}

\subsection{bulk Dirac points}
\label{sec:dirac_points}
We first examine the bulk quasi-energy of the system. Figure \ref{fig:dispersion_homogeneous} shows the quasi-energy $\varepsilon_{\bm{k}}$ of our model for a specific values of $\theta_1$ and $\theta_2$. We find that there are four Dirac points with $\varepsilon=0$ and $\varepsilon=\pi$, respectively. 
The Dirac points with $\varepsilon=0$
are located at $(k_x,k_y)=(X,0),\ (\pi+X,0),\ (-X,\pi)$, and $(\pi-X,\pi)$ with $X=\tan^{-1}[\sin(\theta_2)/\cos(\theta_1)\cos(\theta_2)]$, and those with
$\varepsilon=\pi$ are at  $(k_x,k_y)=(0,\pm\pi/2)$ and $(\pi,\pm\pi/2)$ for arbitrary $\theta_1$ and $\theta_2$.
See Appendix \ref{sec:explicit-form} for details. 

These Dirac points are protected by a (weak) topological number associated with chiral symmetry in Eq. (\ref{eq:chiral}).
To see this, we rewrite $U({\bm k})$ in terms of the Floquet Hamiltonian $H_{\rm F}({\bm k})$,
\begin{align}
U({\bm k})=e^{-iH_{\rm F}({\bm k})}=\cos H_{\rm F}({\bm k})-i\sin H_{\rm F}({\bm k}). 
\label{eq:U_expansion}
\end{align}
Since Eq. (\ref{eq:chiral})  implies the conventional chiral symmetry for $\sin H_{\rm F}({\bm k})$,
\begin{align}
\Gamma \sin H_{\rm F}({\bm k}) \Gamma^{-1}=-\sin H_{\rm F}({\bm k}),
\label{eq:chiral_sin}
\end{align}
one can define the one-dimensional winding number 
\begin{align}
\nu_1:=\frac{i}{4\pi}\oint_C\mathrm{tr}[\Gamma (\sin H_{\rm F}({\mbox{\boldmath $k$}}))^{-1}
d (\sin H_{\rm F}({\mbox{\boldmath $k$}}))],
\label{eq:topological-charge_definition}
\end{align}
where $C$ is a circle enclosing the Dirac point.
Note that $(\sin H_{\rm F}({\bm k}))^{-1}$ is well-defined on $C$ since the quasi-energy on $C$ has a gap at $\varepsilon=0$ and $\pi$.

\begin{figure}[t]
\begin{center}
\includegraphics[scale=0.33]{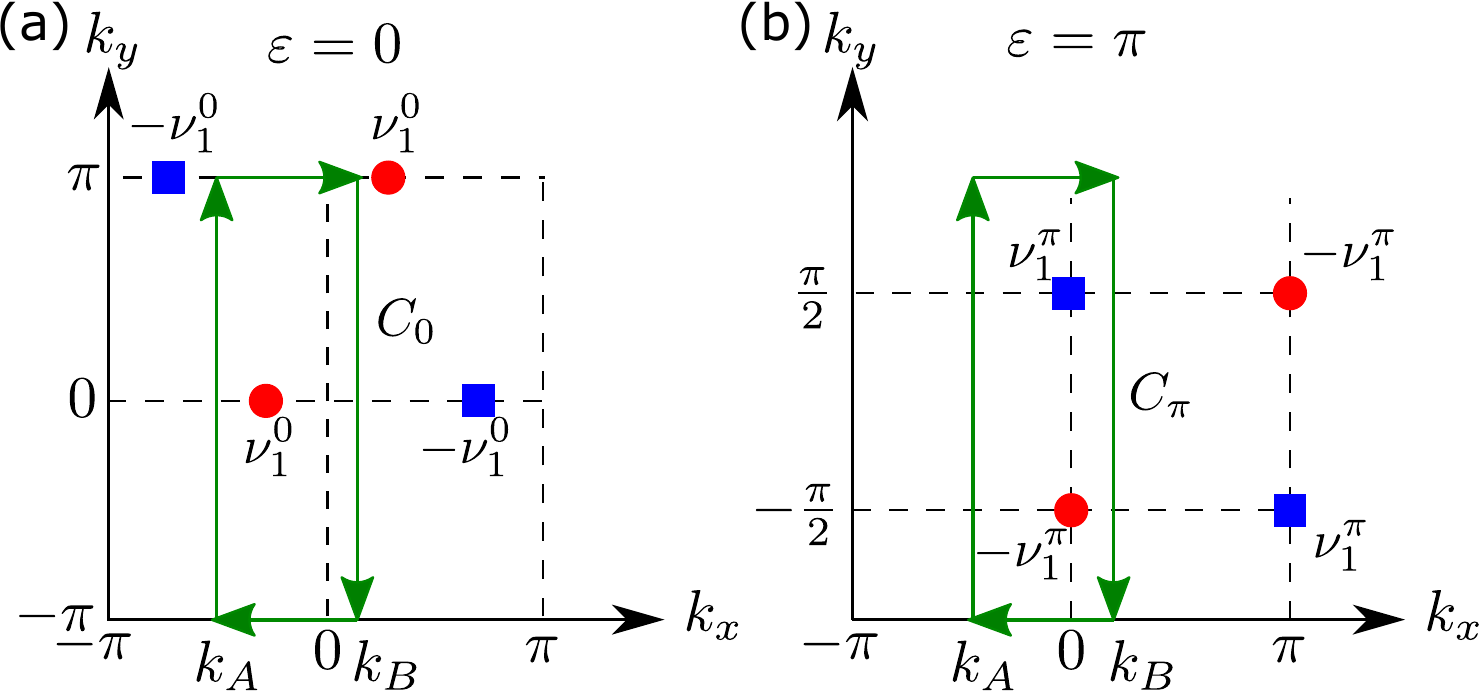}
\caption{The positions and topological charges of Dirac points, (a) for $\varepsilon=0$ and (b) for $\varepsilon=\pi$. In (a), Dirac points are located at $(k_x,k_y)=(X,0),\ (\pi+X,0),\ (-X,\pi)$, and $(\pi-X,\pi)$ where $X$ is defined as  $X=\tan^{-1}[\sin(\theta_2)/\cos(\theta_1)\cos(\theta_2)]$. The figure is described in the case $-\pi/2<X<0$. In (b), the Dirac points are at  $(k_x,k_y)=(0,\pm\pi/2)$ and $(\pi,\pm\pi/2)$.} Dirac points indicated by red circles and blue squares have the opposite topological charges denoted as $\pm\nu_1^0$ and $\pm\nu_1^\pi$ in each figure. Green vertical arrows represent integral paths when we calculate winding numbers fixing $k_x$ at some value, where right ones ($k_x$ is fixed at $k_B$) are directed to the opposite direction to the left ones. Since horizontal arrows cancel out each other due to the $2\pi$ periodicity of $k_y$, the sum of topological charges enclosed by the paths (named as $C_0$ for $\varepsilon=0$ and $C_\pi$ for $\varepsilon=\pi$) corresponds to the difference between winding numbers with $k_x=k_A$ and $k_B$.
\label{fig:dirac-points}
\end{center}
\end{figure}

For the calculation of $\nu_1$, it is convenient to take the basis where $\Gamma$ is diagonal,
\begin{align}
\Gamma=\left(
\begin{array}{cc}
1     &0  \\
0     & -1
\end{array}
\right).
\label{eq:basis_chiral}
\end{align}
In this basis, $U({\bm k})$ has the following form 
\begin{align}
&U({\bm k})=\left(
\begin{array}{cc}
A({\bm k})  & B({\bm k}) \\
-B^{\dagger}({\bm k}) & D({\bm k}) 
\end{array}
\right), 
\nonumber\\
&A({\bm k})=A^{\dagger}({\bm k}), \quad D({\bm k})=D^{\dagger}({\bm k}),
\label{eq:ABD}
\end{align}
because of chiral symmetry (\ref{eq:chiral}). 
Since $\sin H_{\rm F}({\bm k})$ ($\cos H_{\rm F}({\bm k})$) anti-commutes (commutes) with $\Gamma$, the off-diagonal (diagonal) part of $U({\bm k})$ gives $\sin H_{\rm F}({\bm k})$ ($\cos H_{\rm F}({\bm k}))$. Thus, we have
\begin{align}
-i\sin H_{\rm F}({\bm k})=\left(
\begin{array}{cc}
0     & B({\bm k})  \\
-B^{\dagger}({\bm k})     &0 
\end{array}
\right),
\end{align}
from which we obtain
\begin{align}
\nu_1=\frac{1}{2\pi}{\rm Im}\left[\oint_C d\ln \det B({\bm k})\right].    
\end{align}
This topological number is non-zero for each of Dirac points, as summarized in Fig.\ref{fig:dirac-points}.

In terms of $U_1=V_2V_1$, $\nu_1$ is given as follow. 
Denoting $U_1$ in the basis of Eq. (\ref{eq:basis_chiral}) as Eq. (\ref{eq:U1}), from the relation $\sin H_{\rm F}=i(U-U^{\dagger})/2$, we have 
\begin{align}
B({\bm k})=2\alpha^{\dagger}({\bm k})\beta({\bm k})=-2\gamma^{\dagger}({\bm k})\delta({\bm k}).
\label{eq:B}
\end{align}
This equaiton implies that 
\begin{align}
\nu_1=\nu_1[\beta]-\nu_1[\alpha]
=\nu_1[\delta]-\nu_1[\gamma],
\label{eq:topological-charge}
\end{align}
where $\nu_1[\alpha]$, $\nu_1[\beta]$, $\nu_1[\gamma]$, $\nu_1[\delta]$ are the one-dimensional winding numbers of $\alpha$, $\beta$, $\gamma$ and $\delta$ in the contour $C$.
Lemma in Sec. \ref{sec:chiralcase} ensures that $\det\beta$ and $\det\gamma$ ($\det\alpha$ and $\det\delta$) do not become zero when the gap at $\varepsilon=0$ ($\varepsilon=\pi$) is open on  $C$. We find that $\nu_1[\beta]$ and $\nu_1[\gamma]$ ($\nu_1[\alpha]$ and $\nu_1[\delta]$) are non-trivial if the contour $C$ encloses a Dirac point at $\varepsilon=0$ ($\varepsilon=\pi$). Equation (\ref{eq:topological-charge}) results in
\begin{align}
\nu_1=\frac{\nu_1[\beta]-\nu_1[\gamma]}{2}-\frac{\nu_1[\alpha]
-\nu_1[\delta]}{2},
\label{eq:topological-charge_useful}
\end{align}
which we use in Sec. \ref{sec:winding_cs}.

\subsection{winding numbers}
\label{subsec:winding-number}

Since $d$ is even and $\{{\cal G}_{\rm T},\Gamma\}=0$, if there exists a gap at $\varepsilon=\pi$, we can characterize the system by the time-glide chiral winding number, as shown in Table \ref{table:time-glide_chiral}. Also, by fixing $k_x$ or $k_y$ as a parameter, we can regard the two-dimensional system as a one-dimensional chiral symmetric one,
which hosts winding numbers in Table \ref{table:chiral}. The values of winding numbers are listed in Table \ref{tab:winding-number}.
\begin{table}[htbp]
\caption{A Table of various winding numbers studied in Sec. \ref{subsec:winding-number}. The left column represents the wave number which is fixed as a parameter. The first row in the third and fourth columns shows symmetries which are origins of corresponding winding numbers. Note that, while $\nu_1^\text{c}$ in Eq. (\ref{eq:nu1_c}) depends on $k_x$, its value only takes $\nu_1^\text{c}=\pm1$ at $|k_x|=|\pi/2|$ or $\nu_1^\text{c}=0$ at $k_x=0,\pi$.}
\label{tab:winding-number}
\begin{center}
\begin{tabular}{|c|c|c|c|}
\hline
 fixing & gap & chiral symmetry & time-glide chiral symmetry  \\\hline
$k_x$ & $\varepsilon=0$ & $\nu_1^\text{c}=\pm1,\,0$ & undefined \\\hline
$k_x$ & $\varepsilon=\pi$ & zero & undefined \\\hline
$k_y$ & $\varepsilon=0$ & zero & undefined \\\hline
$k_y$ & $\varepsilon=\pi$ & zero & $\nu_1^\text{tgc}=\pm1$ \\\hline
\end{tabular}
\end{center}
\end{table}
\subsubsection{winding numbers from time-glide chiral symmetry}
\label{sec:winding_tgcs}
We calculate winding numbers whose origin is time-glide chiral symmetry, $\nu_1[\alpha^\prime]$, on time-glide chiral symmetric planes, $k_y=0,\,\pi$. When $k_x$ is fixed at some value, there is no winding number from time-glide chiral symmetry, since preserving time-glide chiral symmetry is impossible when we integrate out $k_y$. Fixing $k_y=0$ and integrating out $k_x$, the winding number becomes
\begin{align}
\nu_1^\text{tgc}=\nu_1(\alpha^\prime)=\text{sign}[\sin(\theta_1)\cos(\theta_2)]
\ \text{at}\ k_y=0,
\label{eq:nu1_tgc}
\end{align}
which predicts the number of edge states with $\varepsilon=\pi$. See Appendix \ref{sec:explicit-form} for the explicit form of
$\alpha^\prime(k_x,k_y=0)$. In the case of $k_y=\pi$, the winding number becomes the same as $\nu_1^\text{tgc}$. This relation can be understood by the additional symmetry for this model $U_1'(k_x,k_y+\pi)=U_2'(-k_x,k_y)$. The winding number obeys the following
\begin{align}\nonumber
&\nu_1[\alpha'(k_x,0)]=-\nu_1[\alpha'^\dagger(k_x,0)]\\
&=-\nu_1[\alpha'(-k_x,\pi)]=\nu_1[\alpha'(k_x,\pi)].
\end{align}
In the first equality, we used Eq. (\ref{eq:U2_prime_even_commute}) at ${\bm k}={\bm k_G}$ and a general relation of winding numbers $\nu_1[\alpha^\prime]=-\nu_1[\alpha'^\dagger]$. In the second equality, we used the additional symmetry constraint. In the last equality, we used the fact that the winding number flips when we flip the integral path.
For $\varepsilon=0$, there is no winding number from time-glide chiral symmetry as shown in Table. \ref{table:time-glide_chiral}. Also, as mentioned in Sec. \ref{sec:dirac_points}, for $k_y=0$ and $\pi$, energy gaps close at $\varepsilon=0$ in the present model. Therefore, we cannot define the winding number related to edge states with $\varepsilon=0$ originating from time-glide chiral symmetry at $k_y=0,\,\pi$. The value of $\nu_1^\text{tgc}$ depends on $\theta_1$ and $\theta_2$, as is described in Fig. \ref{fig:winding_number}. We also remark that the winding number originating from time-glide chiral symmetry is not related to Dirac points of the Floquet Hamiltonians in Fig.\ \ref{fig:dirac-points} since time-glide symmetry cannot be defined for the (time independent) Floquet Hamiltonians.

\subsubsection{winding numbers from chiral symmetry}
\label{sec:winding_cs}
\begin{figure}[tbp]
\begin{center}
\includegraphics[scale=0.33]{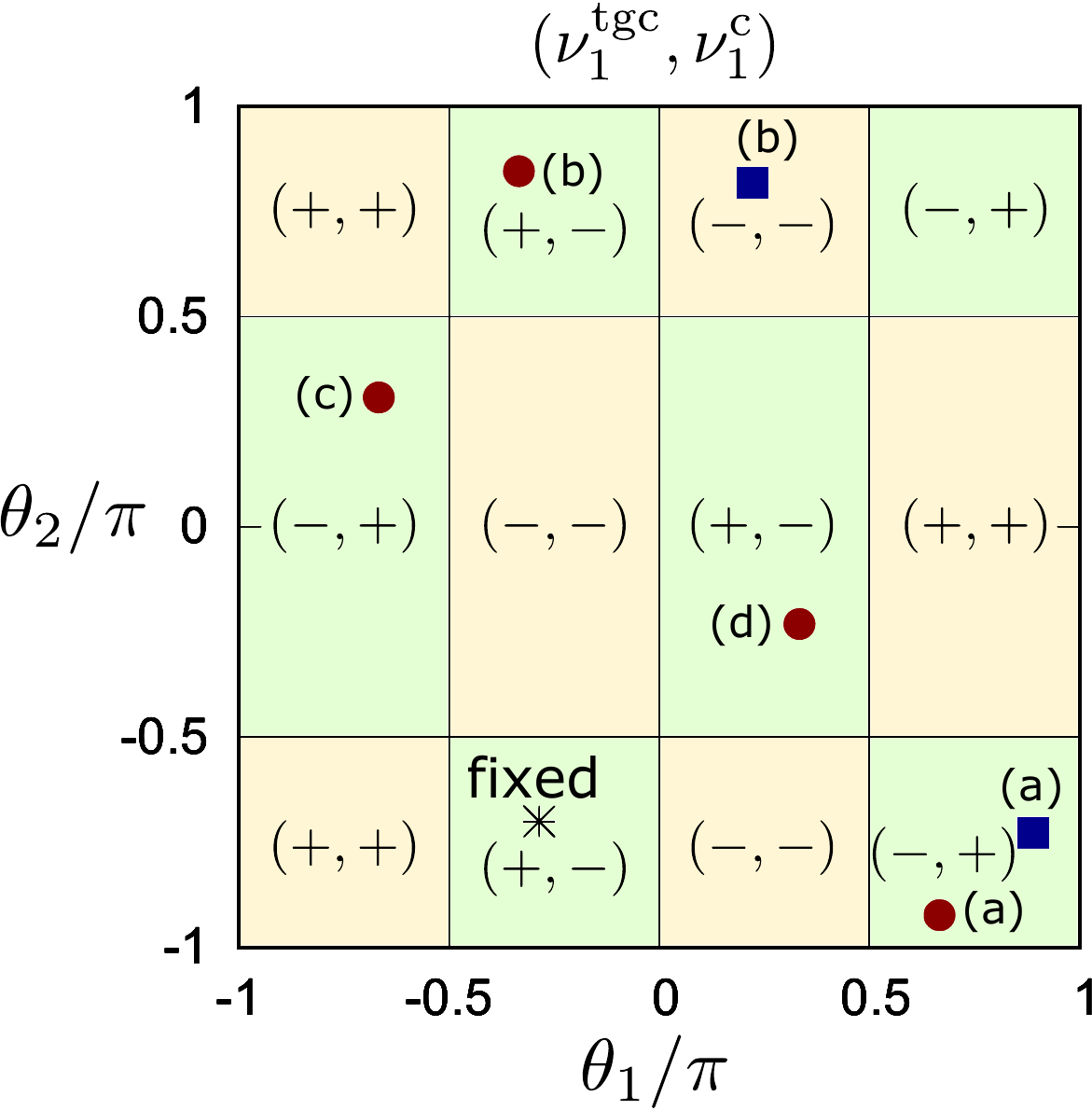}
\caption{$\theta_1$ and $\theta_2$ dependence of winding numbers $(\nu_1^\text{tgc},\nu_1^\text{c})$, defined at $k_y=0$ and $k_x=\pi/2$, respectively. $+1\ (-1)$ is represented as $+\ (-)$. The black asterisk represents fixed parameters ($\theta_1^L,\theta_2^L$) and ($\theta_1^B,\theta_2^B$) in both Fig. \ref{fig:dispersion_x-boundary} and Fig. \ref{fig:dispersion_y-boundary}. Red circles marked (a)-(d) represent parameters ($\theta_1^R,\theta_2^R$) used in Fig. \ref{fig:dispersion_x-boundary}. Blue squares marked (a) and (b) are parameters ($\theta_1^T,\theta_2^T$) used in Fig. \ref{fig:dispersion_y-boundary}.}
\label{fig:winding_number}
\end{center}
\end{figure}

 First, we consider the case of $\varepsilon=0$ fixing $k_x$ as a parameter. Although there are two types of winding numbers, we consider only one type of winding number for each gap. This is because unitary $U_1({\bm k})$ always satisfies $\nu_1[\alpha]-\nu_1[\beta]-\nu_1[\gamma]+\nu_1[\delta]=0$ from Eq. (\ref{eq:unitaryU1}) and $\nu_1[\alpha]+\nu_1[\beta]+\nu_1[\gamma]+\nu_1[\delta]=-4$ is satisfied for arbitrary parameters in the present model. From the above two conditions, two winding numbers in Table. \ref{table:chiral} are related by $\nu_1[\gamma]+1=-\nu_1[\beta]-1$ in our model. Then, we define a winding number for $\varepsilon=0$ as
\begin{align}
\nu_1^\text{c}=\frac{\nu_1[\beta]-\nu_1[\gamma]}{2}
\label{eq:nu1_c}
\end{align}
in the basis where $\Gamma$ becomes $\sigma_3$. $\theta_1,\,\theta_2$, and $k_y$ dependence of $\beta({\bm k})$ and $\gamma({\bm k})$ is written in Appendix \ref{sec:explicit-form}. Although the winding number in Eq. (\ref{eq:nu1_c}) can be calculated at arbitrary $k_x$, it is sufficient to calculate the winding number at $k_x=0,\ \pm\pi/2$, and $\pi$. This is because the value of the winding number changes at Dirac points and Dirac points are located as shown in Fig. \ref{fig:dirac-points} (a). Substituting the specific wave numbers into $\beta({\bm k})$ and $\gamma({\bm k})$ and integrating out $k_y$, the value of $\nu_1^\text{c}$ becomes
\begin{align}
\nu_1^\text{c}&=\mp\text{sign}[\cos(\theta_1)]\ \text{at}\ k_x=\pm\frac{\pi}{2},
\label{eq:nu1_c_pi/2}\\
\nu_1^\text{c}&=0\ \text{at}\ k_x=0,\,\pi.
\label{eq:nu1_c_0}
\end{align}
$\nu_1^\text{c}$ depends on $\theta_1$ as shown in Fig. \ref{fig:winding_number} at $k_x=\pi/2$, while $\nu_1^\text{c}=0$ for arbitrary $\theta_1$ and $\theta_2$ at $k_x=0,\,\pi$. For $\varepsilon=\pi$, we consider a winding number $\frac{\nu_1[\alpha]-\nu_1[\delta]}{2}$ in the same way as the case of $\varepsilon=0$. It is sufficient to consider the winding number at $k_x=\pm\pi/2$, since Dirac points are arranged as shown in Fig. \ref{fig:dirac-points} (b). Then, we find that the winding number becomes zero ($\nu_1[\alpha]$ and $\nu_1[\delta]$ become $-1$) for arbitrary $\theta_1,\,\theta_2$, and $k_x$.\\\indent
Here, we mention the relation between topological charges of Dirac points in Eq. (\ref{eq:topological-charge_useful}) and winding numbers originating from chiral symmetry in Eq. (\ref{eq:nu1_c}), by considering closed integral paths $C_0$ and $C_\pi$ enclosing Dirac points in Fig.\ \ref{fig:dirac-points}. We note that the path $C_\pi$ always enclose even numbers of Dirac points, while the path $C_0$ is not the case. Here, the vertical line with the upward arrow is identical to the integration path for the winding number at $k_x=k_A$, while the vertical line with the downward arrow is the opposite direction of the path for that at $k_x=k_B$. Further, integrals along the two horizontal lines cancel out each other due to $2\pi$ periodicity of $k_y$. Then, the difference between winding numbers with $k_x=k_A$ and $k_x=k_B$, related to edge states at $\varepsilon=0\,(\varepsilon=\pi)$, becomes total topological charges of Dirac points enclosed by $C_0\,(C_\pi)$. This is the reason why the winding number $\nu_1^\text{c}$ is not equal to zero between two Dirac points with opposite topological charges for $\varepsilon=0$ [Fig. \ref{fig:dirac-points} (a)]. In the case of $\varepsilon=\pi$, the total topological charges of Dirac points enclosed by integral paths are always zero, which ensures that the winding number $\frac{\nu_1[\alpha]-\nu_1[\delta]}{2}$ is always zero [Fig. \ref{fig:dirac-points} (b)]. \\\indent
 When $k_y$ is fixed as a parameter, the winding numbers $\frac{\nu_1[\beta]-\nu_1[\gamma]}{2}$ for $\varepsilon=0$ and $\frac{\nu_1[\alpha]-\nu_1[\delta]}{2}$ for $\varepsilon=\pi$ are zero with arbitrary $\theta_1,\,\theta_2$. This behavior of winding numbers can also be understood from topological charges and positions of Dirac points in Fig. \ref{fig:dirac-points}, in the same way as the case of fixing $k_x$.
\begin{figure}[tbp]
\begin{center}
\includegraphics[scale=0.31]{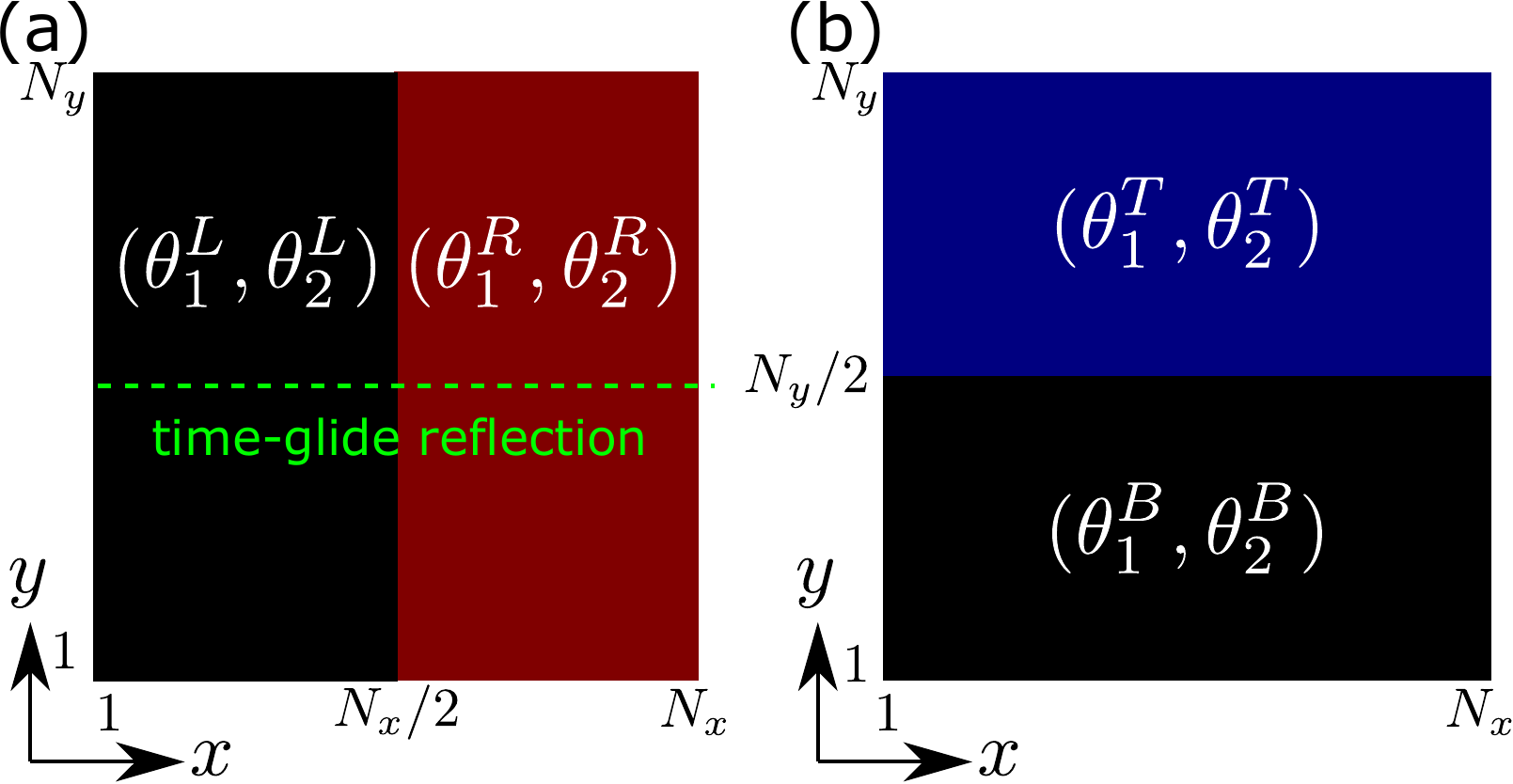}
\caption{Systems with boundaries in the (a) $x$ and (b) $y$ direction. In both cases (a) and (b), periodic boundary conditions are imposed in both directions $x$ and $y$. In (a), where $(\theta_1,\theta_2)$ change at $x=N_x/2$, the green dashed line represents the reflection axis in the $y$ direction by time-glide symmetry. In (b), parameters $(\theta_1,\theta_2)$ change at $y=N_y/2$.}
\label{fig:boundaries}
\end{center}
\end{figure}

\subsection{Edge states}

\label{sec:edge-states}
\begin{figure}[b]
\begin{center}
\includegraphics[scale=0.82]{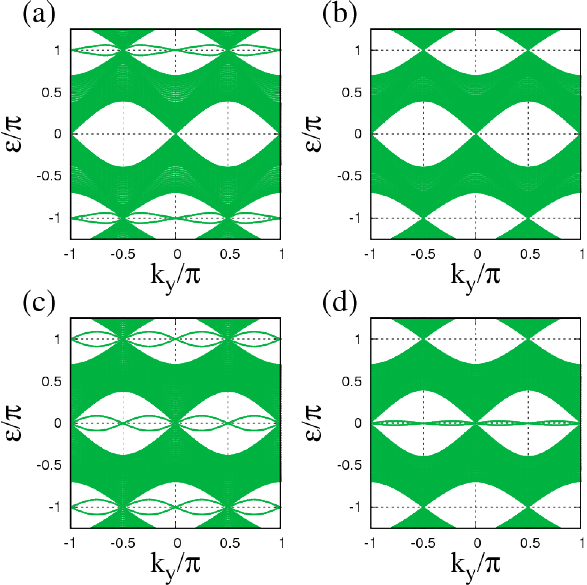}
\caption{Numerically obtained quasi-energies. In (a) and (c), edge
 states with $\varepsilon=\pi$ are protected by time-glide chiral symmetry. In the left region, black region in Fig. \ref{fig:boundaries} (a), $\theta_1^L$ and $\theta_2^L$ are fixed as $\theta_1^L=-2\pi/7,\ \theta_2^L=-7\pi/10$ and the system size is $N_x=160,\ N_y=120$. Periodic boundary conditions are imposed for both $x$ and $y$ direction. Parameters in the right region are (a)\ $\theta_1^R=2\pi/3,\ \theta_2^R=-12\pi/13$,\ (b)\ $\theta_1^R=-\pi/3,\ \theta_2^R=11\pi/13$,\ (c)\ $\theta_1^R=-2\pi/3,\ \theta_2^R=4\pi/13$,\ (d)\ $\theta_1^R=\pi/3,\ \theta_2^R=-3\pi/13$, which are plotted as red circles in Fig. \ref{fig:winding_number}. Note that, there are gap closing points at $(k_y,\varepsilon)=(\pm\pi/2,\pi)$.}
\label{fig:dispersion_x-boundary}
\end{center}
\end{figure}

Here we study the edge states at interfaces between two quantum walks with different parameters $\theta_1, \theta_2$.

\subsubsection{edge states protected by time-glide chiral symmetry}
\label{subsec:time-glide symmetry}
\begin{figure*}[t]
\begin{center}
\includegraphics[scale=0.62]{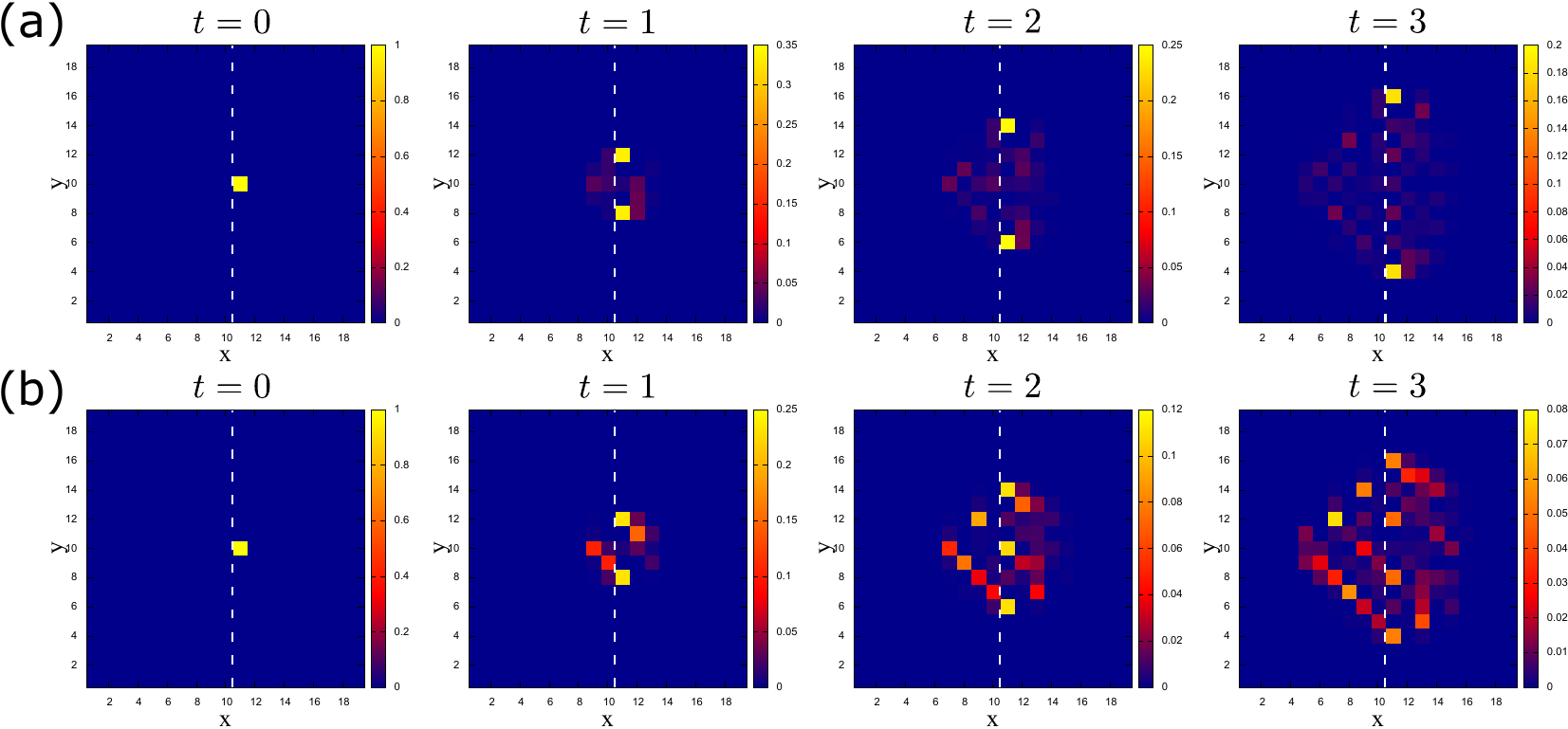}
\caption{The probability distribution of $\ket{\psi(t)}$, $|\psi(x,y,t)|^2=|\psi_+(x,y,t)|^2+|\psi_-(x,y,t)|^2$. In (a) and (b), parameters are the same with Fig. \ref{fig:dispersion_x-boundary} (a) and (b) respectively, that is, edge states with $\varepsilon=\pi$ exist in (a) and there are no edge states in (b). The initial state is $\ket{\psi(0)}=\ket{x=11,y=10,\sigma=-}$ and $N_x=N_y=20$ in both (a) and (b). White dashed lines represent the boundary at which parameters $(\theta_1,\theta_2)$ change.}
\label{fig:time-evolution}
\end{center}
\end{figure*}
Edge states originating from time-glide chiral
symmetry are expected to appear at $(k_y,\varepsilon)=(0,\pi)$ and $(\pi,\pi)$. Making $\theta_1$ and $\theta_2$ position dependent, we can vary $\nu_1^\text{tgc}$ spatially, since the value of $\nu_1^\text{tgc}$ depends on $\theta_1$ and $\theta_2$ as in Fig. \ref{fig:winding_number}. In order to study the edge modes protected by time-glide chiral symmetry, we consider the case where rotation angles depend only on $x$ as described in Fig \ref{fig:boundaries} (a),
\begin{align}
\theta_j(x)=\left\{\begin{array}{l}
\theta_j^L\ (1 \le x \le N_x/2)\\
\theta_j^R\ (N_x/2+1 \le x \le N_x)
\end{array}\right.
,\ j=1,2,
\label{eq:x_dependence}
\end{align}
where an even number $N_x$ is the number of lattices in the $x$
direction. Since periodic boundary conditions are imposed on $x$
and $y$ directions, there are two boundaries at $x=0$ and $N_x/2$
where $\theta_j(x)$ 
change values. \\\indent 
Dispersion relations as a function of $k_y$ are shown in
Fig. \ref{fig:dispersion_x-boundary} for several sets of
$\theta_1^R$ and $\theta_2^R$ by fixing the values of $\theta_1^L$ and
$\theta_2^L$. Comparing the dispersion relation for homogeneous systems
in Fig.\ \ref{fig:dispersion_homogeneous} (c), we can recognize that edge states appear at $(k_y,\varepsilon)=(0,\pi)$ and  $(\pi,\pi)$ in Fig. \ref{fig:dispersion_x-boundary} (a) and (c), while they do not appear in Fig. \ref{fig:dispersion_x-boundary} (b) and (d). 
We confirm that these states are edge states localized near the boundaries. [We note that although edge states originating from extra symmetry appear at $(k_y,\varepsilon)=(\pm\pi/2,0)$ in Fig. \ref{fig:dispersion_x-boundary} (c) and (d), we do not focus on them in this section. See Appendix \ref{sec:additional_symmetry} for details.] In
Fig. \ref{fig:dispersion_x-boundary} (a) and (c), edge states at $(k_y,\varepsilon)=(0,\pi)$ have four-fold degeneracy. We see from Fig.\ \ref{fig:winding_number} that
$|(\nu_1^\text{tgc})_L-(\nu_1^\text{tgc})_R|=2$ in the case of Fig.\ \ref{fig:dispersion_x-boundary}
(a) and (c), where $(\nu_1^\text{tgc})_L$ and $(\nu_1^\text{tgc})_R$ are winding numbers in the left and right regions respectively, while $|(\nu_1^\text{tgc})_L-(\nu_1^\text{tgc})_R|=0$ in the other cases. We remark that the number of edge states predicted by the bulk-edge correspondence becomes twice $|(\nu_1^\text{tgc})_L-(\nu_1^\text{tgc})_R|$, because of the presence of two boundaries at $x=0$
and $N_x/2$. Therefore, $\nu_1^\text{tgc}$ correctly predicts the number of edge states,
in order words, we confirm the bulk-edge correspondence for the
winding number $\nu_1^\text{tgc}$ originating from time-glide chiral symmetry. Accordingly, we clarify that the origin of edge states at $\varepsilon=\pi$ is time-glide
chiral symmetry. It is a theoretically new result that there are edge states at
$\varepsilon=\pi$ protected by time-glide chiral symmetry when
gap closing points exist at $\varepsilon=\pi$, compared with classification in Ref. \cite{morimoto2017floquet} focusing on cases where
bulk spectrum have the energy gap around $\varepsilon=\pi$. When
$k_y\neq0,\pi$, the winding number originating from time-glide
chiral symmetry in Eq. (\ref{eq:time-glide_chiral}) is not defined, and the
edge states at $(k_y,\varepsilon)=(0,\pi)$ deviate from
$\varepsilon=\pi$ away from $k_y=0$ and $\pi$. As a result, the group velocity of edge states is not equal to zero, which is different from that of flat bands from chiral symmetry discussed in Sec. \ref{subsec:chiral symmetry}.\\\indent
Figure \ref{fig:time-evolution} (a) shows the time evolution of
probabilities of a walker when edge states at $\varepsilon=\pi$
exist. We can clearly see that two peaks of probabilities propagate in the opposite directions along the boundary near $x=N_x/2$, which reflects
the existence of edge states with nonzero group velocity. Figure
\ref{fig:time-evolution} (b) shows the time evolution when there are no
edge states. Probabilities diffuse in both $x$ and $y$ directions, and
there are no outstanding peaks in Fig. \ref{fig:time-evolution}
(b). Comparing Fig. \ref{fig:time-evolution} (a) and (b), it is clear
that the propagation of probability peaks along the boundary in
Fig. \ref{fig:time-evolution} (a) is due to the existence of edge
states. Note that, in the dynamics of a 2D quantum walk
with different topological phases studied in
Ref. \cite{chen2018observation}, a peak of probabilities propagates
only to one direction, which is different from the behaviour observed in Fig. \ref{fig:time-evolution} (a).

\subsubsection{edge states protected by chiral symmetry}
\label{subsec:chiral symmetry}
\begin{figure}[tbp]
\begin{center}
\includegraphics[scale=1.1]{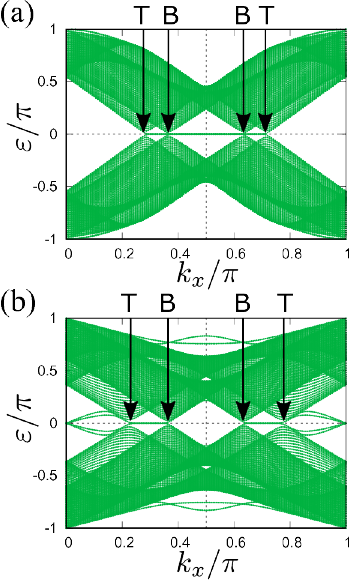}
\caption{Numerically obtained quasi-energies. Flat bands at $\varepsilon=0$ connect Dirac points, corresponding to Fermi arcs of Dirac semimetals. In the bottom region,the  black region in Fig. \ref{fig:boundaries} (b), $\theta_1^B$ and $\theta_2^B$ are fixed as $\theta_1^B=-2\pi/7,\ \theta_2^B=-7\pi/10$ and the system size is $N_x=160,\ N_y=120$. Periodic boundary conditions are imposed for both $x$ and $y$ directions. Parameters in the top region are (a)\ $\theta_1^T=8\pi/9,\ \theta_2^T=-8\pi/11$,\ (b)\ $\theta_1^T=2\pi/9,\ \theta_2^T=9\pi/11$, which are plotted as blue squares in Fig. \ref{fig:winding_number}.}
\label{fig:dispersion_y-boundary}
\end{center}
\end{figure}

Edge states which are protected by chiral symmetry in Eq. (\ref{eq:chiral}) emerge when we make boundaries to $y$ direction,
\begin{align}
\theta_j(y)=\left\{\begin{array}{l}
\theta_j^B\ (1 \le y \le N_y/2)\\
\theta_j^T\ (N_y/2+1 \le y \le N_y)
\end{array}\right.
,\ j=1,2,
\label{eq:y_dependence}
\end{align}
as in Fig. \ref{fig:boundaries} (b). We call the region of $1\le y
\le N_y/2$ and $N_y/2+1 \le y \le N_y$ as the bottom region and top region, respectively.\\\indent
Since the winding number in Eq. (\ref{eq:nu1_c}) remains $\pm 1$ for a finite range of $k_x$, there appear flat bands at $\varepsilon=0$. Figure \ref{fig:dispersion_y-boundary} shows
quasi-energy $\varepsilon$ as a function of $k_x$, in systems shown in Fig. \ref{fig:boundaries} (b). In Fig. \ref{fig:dispersion_y-boundary} (a) and (b), two Dirac points closest to $k_x=\pi/2$ labeled as B originate from the bulk spectrum in the bottom region. The other two Dirac points labeled as T are that of the top region. The system in Fig. \ref{fig:dispersion_y-boundary} (a) has $\nu_1^\text{c}=-1$ in the bottom region and $\nu_1^\text{c}=+1$ in the top region, as shown in Fig.\ \ref{fig:winding_number}. As predicted from the bulk-edge correspondence, there appear flat bands at $\varepsilon=0$ between Dirac points. The eigenstates with $\varepsilon=0$ between two Dirac points B and B have four-fold degeneracy, which is also consistent with the bulk-edge correspondence since there are two boundaries. As we explained in Sec.\ \ref{sec:winding_cs}, the winding numbers at $k_x=0,\pi$ are zero. Varying $k_x$ from $\pi/2$ to $0$ or $\pi$ and passing through one nearest Dirac point B, the winding number of the bottom region changes from $-1$ to $0$. Then, the $\varepsilon=0$ degeneracy at the specific $k_x$ becomes $2$, since the winding number in the top region is still $-1$. After a second Dirac point T is passed, the flat band vanishes, since winding numbers are zero in both regions. In Fig. \ref{fig:dispersion_y-boundary} (b), $\nu_1^\text{c}=-1$ in both regions. Therefore, flat bands cancel out around $k_x=\pi/2$, between two Dirac points labeled as B. On the other hand, there are flat bands in the $k_x$ range where the winding number is $0$ in the bottom region and $-1$ in the top region, between two Dirac points B and T, one is that of the bottom region and the other is that of the top region. The flat bands in Fig. \ref{fig:dispersion_y-boundary} correspond to Fermi arcs in Dirac semimetals, since the behavior of winding numbers can be understood from topological charges of Dirac points, as explained in Sec. \ref{sec:winding_cs}. While flat bands have been discussed in quantum walks \cite{endo2017sensitivity}, the correspondence to Fermi arcs is firstly stated in this work by clarifying the relation between Dirac points in the bulk and flat bands on the edge. In Fig. \ref{fig:dispersion_y-boundary} (b), there also appear edge states at $(k_x,\varepsilon)=(0,0)$ and $(\pi,0)$. We explain the origin of them in Appendix \ref{sec:additional_symmetry}.

\section{summary}
\label{sec:summary}
We have studied Floquet topological phases of discrete quantum walks with time-glide symmetry, chiral symmetry, or both of them. Identifying chiral symmetry as a discrete version of time-reflection symmetry and using the concept of asymmetric unit in time direction, we have shown the way of constructing models which have these symmetries. Based on the discrete space-time symmetries, we have clarified topological numbers which characterize anomalous topological phases intrinsic to Floquet systems. Topological numbers which we have revealed include ones which are not mentioned in Ref. \cite{morimoto2017floquet}. This is because discrete quantum walks without microscopic Hamiltonians are topologically distinct from ordinary Floquet systems which are described by time-dependent Hamiltonians. Our first comprehensive study on space-time symmetries and resulting topological phases in discrete quantum walks not only has theoretical novelty but also help observations of edge states peculiar to discrete Floquet systems, whose origin is these symmetries, since quantum walks are feasible experimental platforms to explore Floquet topological phases. \\\indent
Using the asymmetric unit construction, we have made a model of two-dimensional quantum walks, which satisfies time-glide symmetry and chiral symmetry. We have calculated winding numbers and have shown that there are two types of edge states. One type of edge states is protected by time-glide chiral symmetry, which appears when there are boundaries only in $x$ direction. Since the edge states have nonzero group velocities, two peaks of probabilities propagate in two opposite directions along the boundary, which can be observed in experiments. While Ref. \cite{morimoto2017floquet} classifies Floquet topological phases when bulk spectrum are fully gapped around $\varepsilon=\pi$, in our model, robust edge states protected by time-glide chiral symmetry appear even when bulk gaps are closed at $\varepsilon=\pi$. The other type of edge states is the flat band, appearing when boundaries exist in $y$ direction and time-glide symmetry is broken, while chiral symmetry is preserved. Existence or absence of flat bands are understood from topological charges of Dirac points. The flat bands correspond to Fermi arcs in Dirac semimetals since the way of understanding is the same with Fermi arcs as explained in Sec. \ref{sec:winding_cs}, which is first stated in quantum walks. While a linear combination of chiral winding numbers in Table. \ref{table:chiral} and a time-glide chiral winding number in Table. \ref{table:time-glide_chiral} lead to the existence of edge states in our model, it should be interesting to explore phenomena related to other topological numbers in discrete quantum walks, such as linear combinations of chiral winding numbers different from Eq. (\ref{eq:nu1_c}), time-glide chiral winding numbers when ${\cal G}_{\rm T}$ and $\Gamma$ commute, or time-glide Chern numbers in systems with odd $d$, as future works.
\section*{acknowledgement}
K. M and T. B thank Ken Shiozaki for suggesting exploration of time-glide symmetry in quantum walks. K. M and H. O thank Yasuhiro Asano and Kousuke Yakubo for helpful discussions. This work was supported by KAKENHI (Grants No. 17H02922, No. JP18J20727, No. JP18H01140, No. JP18K18733, and No. JP19K03646), a Grant-in-Aid for Scientific Research on Innovative Areas (KAKENHI Grant No. JP15H05855 and No. JP18H04210) from the Japan Society for the Promotion of Science, and CREST Grant No. JPMJCR19T2, from Japan Science and Technological Agency. K. M was supported by Atoms, a visiting program of Yukawa Institute for Theoretical Physics in Kyoto University.
\appendix
\setcounter{equation}{0}
\setcounter{table}{0}
\setcounter{figure}{0}
\section{matrix components of $U({\bm k})$, $U_1^\prime({\bm k})$, and $U_1({\bm k})$ for obtaining $\varepsilon({\bm k})$, $\nu_1^\text{tgc}$, and $\nu_1^\text{c}$}
\label{sec:explicit-form}
From Eqs. (\ref{eq:V1_k})-(\ref{eq:V4}), the matrix components of the two by two time-evolution operator $U(\mbox{\boldmath $k$})=V_4(\mbox{\boldmath $k$})V_3(\mbox{\boldmath $k$})V_2(\mbox{\boldmath $k$})V_1(\mbox{\boldmath $k$})$ are
\begin{align}
U^{(1,1)}(\mbox{\boldmath $k$})&=(a+b+c)^2-(d+e+if)^2,
\label{eq:U11}\\
U^{(1,2)}(\mbox{\boldmath $k$})&=2(a+b)(d+e)+2cf,
\label{eq:U12}\\
U^{(2,2)}(\mbox{\boldmath $k$})
&=[U^{(1,1)}(\mbox{\boldmath $k$})]^\ast,\ 
U^{(2,1)}(\mbox{\boldmath $k$})
=-U^{(1,2)}(\mbox{\boldmath $k$})
\label{eq:U22_U21},
\end{align}
where $a,\,b,\,c,\,d,\,e,\,f$ are defined as
\begin{align}
a&=\cos(\theta_1)\cos(\theta_2)\cos(k_x)\cos(k_y),
\label{eq:small-a}\\
b&=\sin(\theta_2)\sin(k_x),
\label{eq:small-b}\\
c&=\cos(\theta_2)\sin(k_y),
\label{eq:small-c}\\
d&=\cos(\theta_1)\cos(\theta_2)\sin(k_x)\cos(k_y),
\label{eq:small-d}\\
e&=-\sin(\theta_2)\cos(k_x),
\label{eq:small-e}\\
f&=\sin(\theta_1)\cos(\theta_2)\cos(k_y).
\label{eq:small-f}
\end{align}
Here, $U^{(i,j)}(\mbox{\boldmath $k$})$ denotes the $(i,j)$ component of $U(\mbox{\boldmath $k$})$. Due to the structure of $U({\bm k})$ in Eq. (\ref{eq:U22_U21}) and $\det(U(\bm k))=1$, the condition for $\varepsilon_{\bm k}=0$ is $U^{(1,1)}(\mbox{\boldmath $k$})=U^{(2,2)}(\mbox{\boldmath $k$})=1$. Then, substituting $k_y=0$ into Eqs. (\ref{eq:U11})-(\ref{eq:small-f}), the condition becomes
\begin{align}
\sin(k_x-X)=0,\ \tan(X)=\frac{\sin(\theta_2)}{\cos(\theta_1)\cos(\theta_2)}.
\label{eq:condition}
\end{align}
Therefore, gap closing points always exist at $(k_x,k_y,\varepsilon)=(X,0,0)$ and $(\pi+X,0,0)$. In the case of $k_y=\pi$, there are also gap closing points at $(-X,\pi,0)$ and $(\pi-X,\pi,0)$, which is derived in the same way.\\\indent
 In order to obtain winding numbers, we also need matrix components of $U_1^\prime({\bm k})=V_1({\bm k})V_4({\bm k})$ and $U_1({\bm k})=V_2({\bm k})V_1({\bm k})$. Calculating $\nu_1^\text{tgc}$ in Eq. (\ref{eq:nu1_tgc}), we integrate $\alpha^\prime(k_x,k_y=0)$, which becomes
\begin{align}
\alpha^\prime(k_x,0)&=a\cos(k_x-b)+ic\sin(k_x),
\label{eq:alpha_prime}\\
a&=\sqrt{\cos^2(\theta_2)+\cos^2(\theta_1)\sin^2(\theta_2)},
\label{eq:a}\\ 
\tan(b)&=\frac{\cos(\theta_1)\sin(\theta_2)}{\cos(\theta_2)},\ c=\sin(\theta_1)
\label{eq:bc}
\end{align}
The above equations tell us that the value of $\nu_1^\text{tgc}$ depends on the sign of $\cos(\theta_2)$ and $\sin(\theta_1)$. Calculating $\nu_1^\text{c}$ in Eq. (\ref{eq:nu1_c}), we change the basis to make chiral operator $\sigma_3$, {\it i.e.} apply a unitary transformation with $e^{\frac{\pi}{4}\sigma_2}$. Then,  from Eqs. (\ref{eq:V1_k}) and (\ref{eq:V2}), $\beta({\bm k})$ and $\gamma({\bm k})$ become 
\begin{align}
\beta({\bm k})&=e^{-ik_y}\cos(\theta_2)[\cos(\theta_1)\cos(k_y)-i\sin(k_y)],
\label{eq:beta}\\
\gamma({\bm k})&=e^{-ik_y}\cos(\theta_2)[\cos(\theta_1)\cos(k_y)+i\sin(k_y)],
\label{eq:gamma}
\end{align}
at $k_x=\pi/2$, which results in $\theta_1$ dependence of $\nu_1^\text{c}$. 

\section{additional symmetries and resulting winding numbers}
\label{sec:additional_symmetry}
The system has an additional symmetry at $k_y=\pm\pi/2$. For simplicity, we focus only on $k_y=\pi/2$. Substituting $k_y=\pi/2$ into $U^\prime({\bm k})$, we obtain
\begin{align}
U^\prime(k_x,\frac{\pi}{2})&
=-\tilde{U}_2(k_x)\tilde{U}_1(k_x),
\label{eq:ky=pi/2}\\
\tilde{U}_1(k_x)
&=C_1(\frac{\theta_2}{2})Z(k_x)C_1(-\frac{\theta_2}{2}),
\label{eq:U_tilde_former}\\
\tilde{U}_2(k_x)
&=C_1(-\frac{\theta_2}{2})Z(-k_x)C_1(\frac{\theta_2}{2}),
\label{eq:U_tilde_latter}
\end{align}
with a unitary transformation by $e^{i(\frac{\theta_1}{2}-\frac{\pi}{4})\sigma_1}e^{i\frac{\pi}{4}\sigma_3}$. $\tilde{U}_1(k_x)$ and $\tilde{U}_2(k_x)$ satisfy
\begin{align}
\sigma_3\tilde{U}_1(k_x)\sigma_3
=\tilde{U}_2^\dagger(k_x).
\label{eq:additional-symmetry_ky=pi/2}
\end{align}
Based on the additional symmetry in Eq. (\ref{eq:additional-symmetry_ky=pi/2}), we can obtain a winding number of
\begin{align}
[\tilde{U}_1(k_x)]^{(1,1)}=\cos(k_x)+i\cos(\theta_2)\sin(k_x),
\label{eq:U1_tilde_11}
\end{align}
which corresponds to a winding number for $\varepsilon=\pi$ in Table \ref{table:chiral}. At $k_y=-\pi/2$, the winding number has the opposite sign to that at $k_y=\pi/2$. The origin of edge states at $(k_y,\varepsilon)=(\pm\pi/2,0)$ in Fig. \ref{fig:dispersion_x-boundary} (b) and (c) is the symmetry in Eq. (\ref{eq:additional-symmetry_ky=pi/2}) or the resulting winding number, since there is the minus sign in the right hand side of Eq. (\ref{eq:ky=pi/2}). At $k_y=\pm\pi/2$, energy gap always closes at $\varepsilon=\pi$. Therefore, we cannot define the winding number related to edge states with $\varepsilon=\pi$ at $k_y=\pm\pi/2$.

The system also has an additional symmetry at $k_x=0,\ \pi$. Substituting $k_x=0$ and ignoring $e^{\pm ik_y}$ which cancel out, $U_1^\prime(0,k_y)$ and $U_2^\prime(0,k_y)$ are the same operator,
\begin{align}
U_1^\prime(0,k_y)=U_2^\prime(0,k_y)
=C_2(-\frac{\theta_2}{2})e^{-ik_y\sigma_1}C_2(-\frac{\theta_2}{2}),
\label{eq:kx=0}
\end{align}
with a unitary transformation by $e^{i\frac{\pi}{4}\sigma_2}$. Since $U_1^\prime(0,k_y)$ satisfies
\begin{align}
\sigma_3U_1^\prime(0,k_y)\sigma_3=[U_1^\prime(0,k_y)]^\dagger,
\label{eq:additional-symmetry_kx=0}
\end{align}
we can obtain a winding number of 
\begin{align}
[U_1^\prime(0,k_y)]^{(1,2)}
=-\sin(\theta_2)\cos(k_y)-i\sin(k_y),
\label{eq:U1_prime_12}
\end{align}
which is a winding number in Table \ref{table:chiral} for $\varepsilon=0$. Edge states at $(k_x,\varepsilon)=(0,0)$ in Fig. \ref{fig:dispersion_y-boundary} (b) originate from the winding number of $[U_1^\prime(0,k_y)]^{(1,2)}$ in Eq. (\ref{eq:U1_prime_12}). At $k_x=\pi$, there appear edge states in the same way.

\bibliographystyle{apsrev4-1}
\bibliography{reference.bib}

\end{document}